\documentstyle[aps,preprint,epsf]{revtex} 
\tightenlines

\begin{document} 
\draft
%\twocolumn[\hsize\textwidth\columnwidth\hsize\csname
%@twocolumnfalse\endcsname
\title{Adsorption of mono- and multivalent cat- and anions on DNA
  molecules.}
\author{E. Allahyarov$^{1}$,\,\, H. L\"owen$^{2}$\, and \, G. 
Gompper$^{1}$ }
\address{{1} Institute\, f\"ur\, Festk\"orperforschung, Forschungszentrum 
J\"ulich, \,\mbox{D-52425} \, J\"ulich, Germany }
\address{{2} Institut\, f\"{u}r\, Theoretische \,Physik
 II,\,Heinrich-Heine-Universit\"{a}t\,
 D\"{u}sseldorf,\,\mbox{D-40225}\,D\"{u}sseldorf, \,Germany}

\date{\today}

\maketitle
\begin{abstract}
Adsorption of monovalent and multivalent cat- and
anions on a deoxyribose nucleic acid (DNA) molecule
 from a salt solution is investigated by
computer simulation.
The ions are modelled as charged hard spheres,
the DNA molecule as a point charge pattern
following the double-helical phosphate strands.
The geometrical shape of the DNA molecules
is modelled on different levels ranging from a simple
cylindrical shape to structured models which
include the major and minor grooves between the phosphate
strands. The  densities of the ions adsorbed
on the phosphate strands, in the major and in the minor grooves
are calculated. First, we find that the adsorption pattern on the DNA surface
depends strongly on its geometrical shape:
counterions adsorb preferentially along the phosphate strands
for a cylindrical model shape, but in the minor groove
for a geometrically structured model. 
Second, we find that an addition of monovalent salt
ions results in an increase of the charge density in the
minor groove while the total charge density of
ions adsorbed in the major groove stays unchanged.
The adsorbed ion densities are highly structured along the 
minor groove while they are almost smeared along the major groove.
Furthermore, for a fixed amount of added salt,
the major groove cationic charge is independent on the counterion
valency. For increasing salt concentration
the major groove is neutralized while the total
charge adsorbed in the minor groove is constant. 
DNA overcharging is detected for
multivalent salt. Simulations for a larger ion radii, which mimic the
effect of the ion hydration, indicate an increased adsorbtion of
cations in the major groove.    

\end{abstract}
\pacs{PACS: 87.15.Aa, 61.20.Ja, 82.70.Dd, 87.10.+e}
%]
%\renewcommand{\thepage}{\hskip 8.9cm \arabic{page} \hfill Typeset
%using REV\TeX }
%\narrowtext
\section{Introduction}

In addition to its biological role as the carrier of genetic
information \cite{saenger}, deoxyribose nucleic acid 
 (DNA) in solution exhibits typical polyelectrolyte
behavior. Its physico-chemical properties, such as
melting temperature, transition between different DNA forms, binding
interaction of proteins and other ligands, strongly depend on the
added salt concentration in aqueous solution. 
Some physical properties of DNA, such as its
osmotic pressure and activity coefficients \cite{katchalsky},
can be explained within a simple polyelectrolyte picture of DNA
as obtained via  counterion condensation
theory of Manning \cite{manning1}, Poisson-Boltzmann (PB) theory
\cite{mills,podgornik2002pre,marcus1955,1951}
and Monte Carlo simulations 
\cite{mills,murthy,mills1986,vlachy,paulsen1987,paulsen1988}
of a homogeneously charged cylinder. 
Although a cylinder with an effective  homogeneous line charge density 
might be
an appropriate model for studying properties far away from the DNA 
surface,
more details are getting relevant if one approaches the DNA surface.
These details can be classified in the following way: first the actual
{\it charge distribution} along the DNA molecule is not a homogeneous line 
charge
but a discrete set of phosphate charge groups  
along a double helix. Next, the {\it geometric shape} of the DNA including 
the major
and minor grooves between the two phosphate strands is getting relevant.
Finally, {\it molecular details} of the DNA surface should be considered 
including
specific interactions and explicit solvent molecules.

The PB  calculations have been successfully applied to
investigate the DNA
electrostatics, such as the electrostatic field of double-helix
charge distribution 
\cite{lin1995} and of all-atom DNA models \cite{fogolari1999,hecht1995}.
The comparison between the simulation and PB results for the multivalent
counterion 
distribution around DNA, addressed in 
Refs.\cite{mills,vlachy,hecht1995,zakharova1999,das1995jcm}, reveals great
differences between them. The reason for
such differences is the lack of the hard core and Coulomb correlations in
mean-field based  PB theory. On the other hand, an all-atom  DNA
simulations in solution
is computationally very costly and 
can be performed  only with small system sizes and low salt concentrations
\cite{feig1998,feig1999,olmsted1989,olmsted1995,allison1994,yang1995bj,lamm1997qch,jaya1996annu,Jayapura1989,young2,gurlie1995}.  
Therefore we focus in this paper on a ``primitive
approach'' \cite{Hansen_rev} which goes beyond the simplistic model of a 
line-charged cylinder
but still does not include full atomic details. The physical reason to do 
so
is that most of the general properties of DNA are expected to result from
a combination of Coulomb and excluded volume interactions which are the 
most
dominating parts of the total interactions for long and short ranges 
\cite{bonvin2000ebj}. Following
this strategy, we disregard the discrete structure of water replacing 
it
by a dielectric background, but treat the salt ions, the double-helical
charge pattern on the DNA molecule and the geometrical grooved shape of
the DNA molecule explicitly.

This paper focuses on the cat- and anion adsorption pattern on the DNA
surface. In particular, in contrast
to earlier simulations which only present rotationally averaged
data for adsorbed ions 
\cite{feig1999,lyubar,korolev1999,montoro1995,montoro1998,montoro2001,gulbrand1989},
 we resolve the adsorption in the major and  minor
grooves and on the phosphate strands. Both a qualitative and a 
quantitative
knowledge of the adsorption pattern is desirable, since it
is needed as a crucial input in other more coarse-grained approaches like 
the Kornyshev-Leikin (KL) theory of DNA-DNA
interaction \cite{kornyshevandleikin}. It is known that the details
of the adsorption pattern strongly influence the effective interaction
forces and even governing the sign of the interaction. Hence the
adsorption pattern will have direct consequences for aggregation
and bundling of DNA molecules caused by an effective mutual
attraction \cite{netz2001epje,deng2000,holm2002macromolecules}.
It is also known that adding multivalent ions
to the solution causes drastic changes in the ion adsorption and in
the DNA aggregation
and bundling \cite{lyubartsev1995,Deng,deng1999}. 
Many experimental facts have been collected regarding the adsorption
of multivalent ions on the DNA. To name just a few,
 there is experimental evidence
 indicating that Mn$_{2+}$ and Cd$_{2+}$ condense on DNA 
\cite{knoll1988book,rau1992biophys}
 but Ca$_{2+}$ or Mg$_{2+}$ do not 
\cite{rau1992biophys,bloomfield1996cosb}.
This points to an important specificity 
\cite{korolev,korolev2001biopolymers,korolev2002biophys,sponer2002,duguid1993} 
and the question arising is whether
this can be understood in simple terms of effective ion radii.
Multivalent counterions of valency larger than 2, on the other hand, 
such as trivalent spermidine (Spd) and
tetravalent spermine
(Spm) are believed to play a key role in
maintaining cellular DNA in a compact state
\cite{cohen1998book,tabor1984,drew1981,feuerstein}. The compactification
of DNA \cite{deng2000,holm2002macromolecules} seems to be  mediated by 
their adsorption
on the DNA surface \cite{deng2000,gosule1976and1978,wilson1}.
Therefore there is a need to study the role of
high-valency counterions in the DNA adsorption pattern
in a systematic way.
 
While the  Manning condensation theory on a homogeneously charged
cylinder is well-studied by now \cite{lebretzimm}, it is a priori
unclear how the adsorbed counterions
will partition themselves in the two grooves and on the phosphate strands. 
Moreover,
the adsorption of anions, which carry  the same charge as the phosphates, 
in the major
grooves is an interesting issue, in particular for higher counterion 
valencies.

The goal of this paper is twofold: first, we aim to predict both 
qualitatively
and quantitatively the nature of the adsorption pattern on a single
DNA molecule for a given added salt concentration, given microion 
valencies
and microion radii. Ion-specific effects, however, only enter via the ion 
sphere
radius and its charge. Although the actual numbers may be influenced by
further details such as the dielectric properties of water at close ion 
distances,
we think that the {\it trends} of our findings upon a change in ion 
valency
and salt concentration will be robust. In particular for high valency
ions, the interactions should be dominated by the Coulombic part such
that our ``primitive'' model should be more appropriate.
Second, on a more technical level, we would like to investigate 
the influence of the geometrical structure of the DNA shape used in the 
theoretical model
of the adsorption pattern. It is expected that a grooved shape will 
attract
more counterions into the grooves electrostatically such that the 
adsorption pattern
will depend  on the geometrical shape used in the model.

In our computer simulations we find that the adsorption pattern on the DNA 
surface
depends strongly on the geometrical model shape of the DNA surface.
In detail three different shapes, modelling the geometrical structure of 
the grooves
on different levels, are considered. It is found that
counterions adsorb preferentially along the phosphate strands
for a cylindrical model shape but in the minor groove
for a geometrically structured model.

Furthermore, we find that an addition of more monovalent salt
ions, provided the counterions are also monovalent, results in an
increase of the charge density in the
minor groove, while the total charge density of
ions adsorbed in the major groove stays unchanged.
The adsorbed ion densities are highly structured along the 
minor groove while they are almost smeared along the major groove.

We also analyze the influence of the ion valency on the ion adsorption
pattern on the DNA surface. We show that 
for any fixed amount of salt the major groove cationic charge is
constant for any counterion valency. For
added multivalent salt, we show the existence of a major groove
neutralization phenomena while the minor groove total charge remains constant.  

We also address the DNA overcharging phenomenon, which is 
of special interest in biology, for example, for
the delivery of genes to the living cell for the purpose of the gene
therapy. Since both the bare DNA and cell surface are
negatively charged, normally DNA does not approach
cell surface.  An overscreened DNA molecules, however,
is effectively positively charged such that it
 could pass through the negative cell membrane. 
Our simulations show that the overcharging of DNA appears generally in
multivalent salt solution regardless the counterion valency. 
Finally we have performed a few simulations with larger ion radii in order to mimic
a larger  ion hydration shell in the solvent. Our findings show that
for an increasing ion radius more cations go to the
major groove
whereas the minor groove and strand ionic occupations shrink.     
 
The  paper is organized as follows: 
In section II we discuss different models for the
DNA shape. Our simulation technique and
model parameters are presented in section III.
Results are given in section IV. Finally we
conclude in section V.

\section{Different models for the DNA shape}
We consider the B-form of DNA which is the most common state of
DNA in aqueous solutions. It has 
an inner core formed by nucleotide pairs, and two
sugar-phosphate strands spiraling around the core. The latter forms a 
well-known double helix with a pitch length of about 34\AA \, and a core radius of 
about 9\AA. There are two phosphate groups per base pair, and 10 base pairs per
pitch (or helical turn). The axial rise per base pair along the DNA
long axis is 3.4\AA \,, hence there is one elementary charge per each
1.7\AA. The average value of the angle between the adjacent
base pairs is $36^0$ which makes the average distance between neighboring
charges on the DNA surface to be 7\AA. This distance, which is much smaller 
than
the helical pitch, is  of the order of the Debye screening length under
physical conditions. Finally the helix persistence length is about 500\AA. 

\begin{figure}
\hspace{4cm}
   \epsfxsize=7cm
   \epsfysize=7.cm 
 ~\hfill\epsfbox{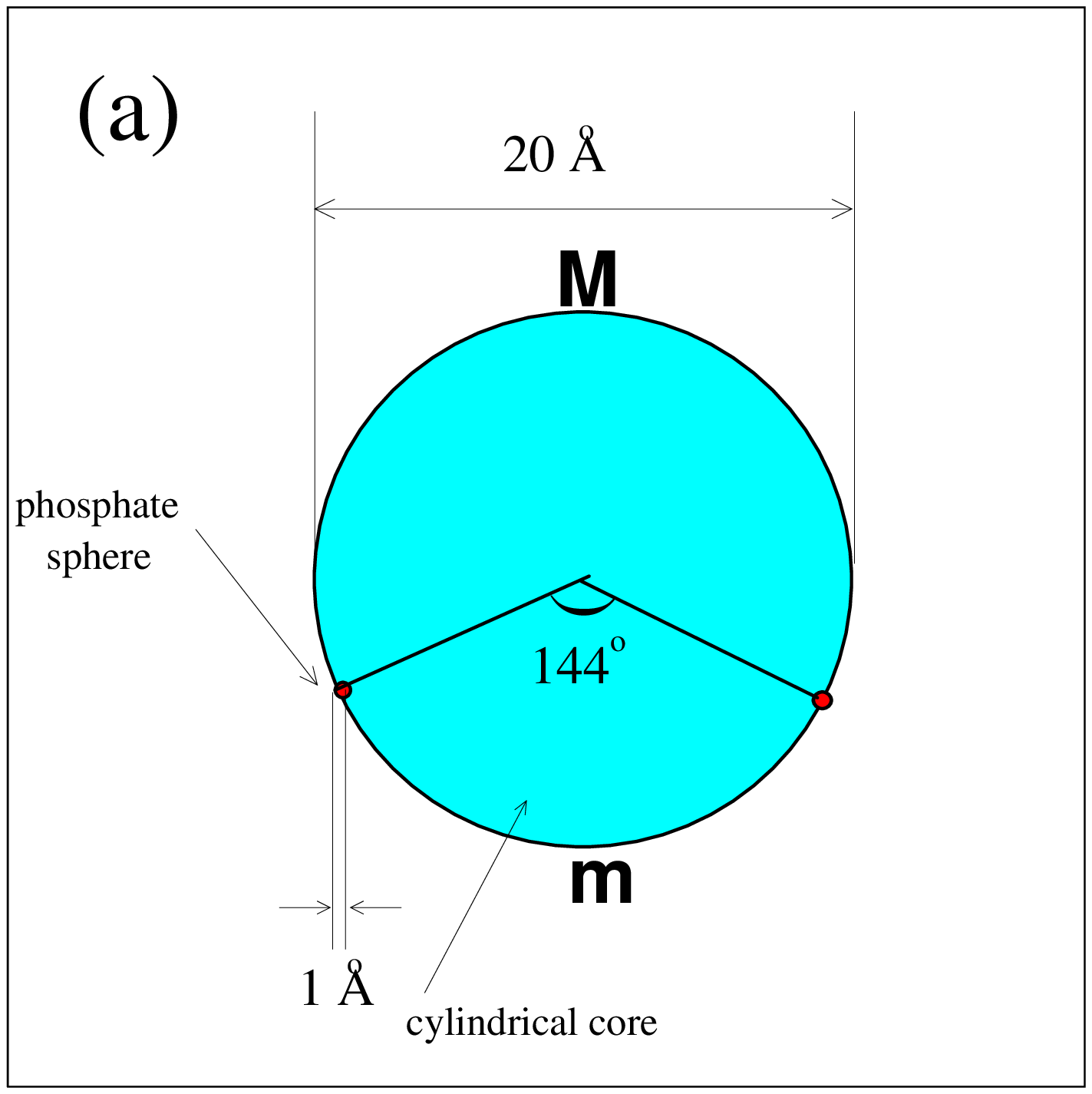}
    \epsfxsize=7cm 
    \epsfysize=7.cm
 ~\hfill\epsfbox{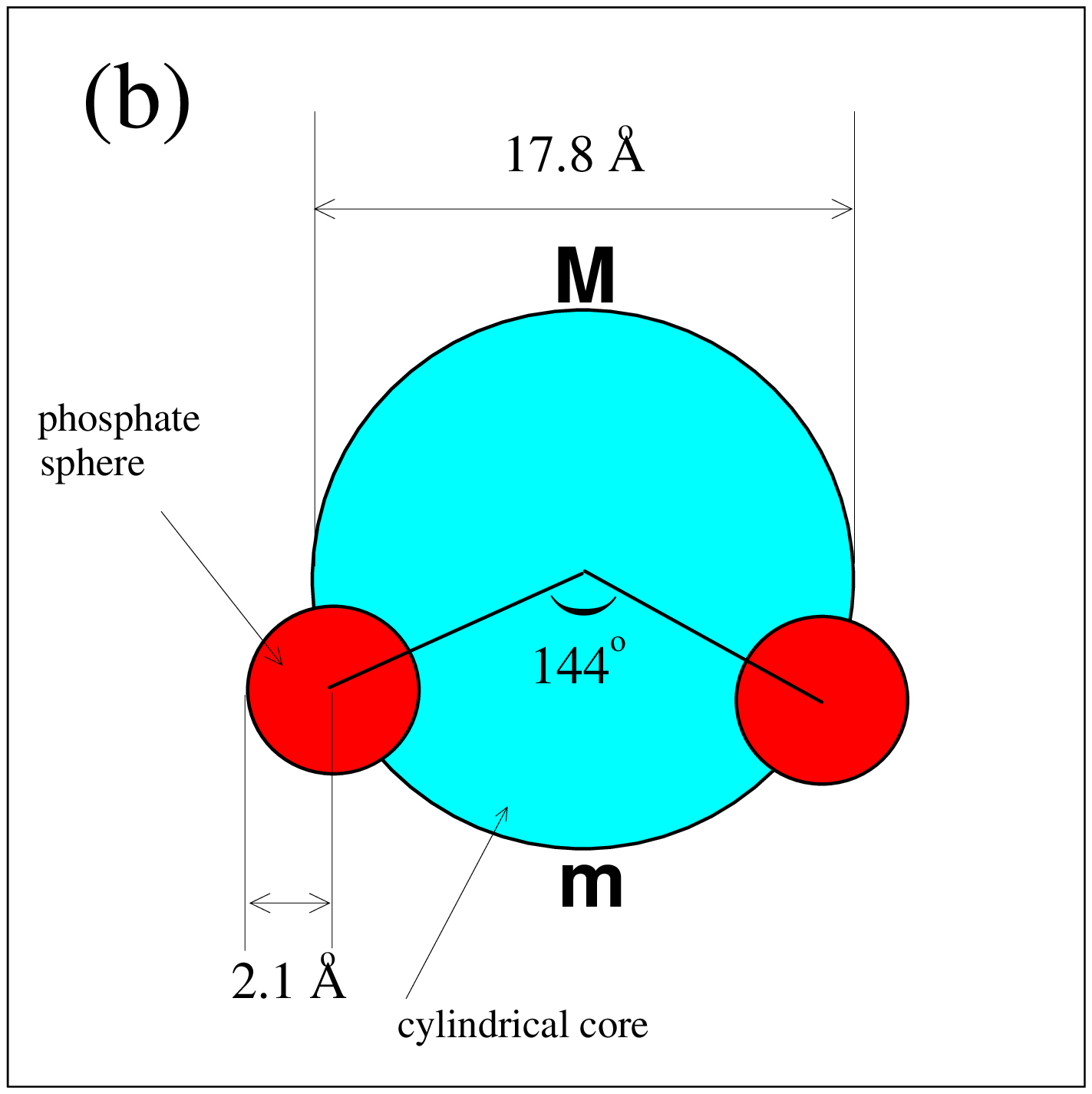}\hfill~
    \epsfxsize=7cm 
    \epsfysize=7.cm
 ~\hfill\epsfbox{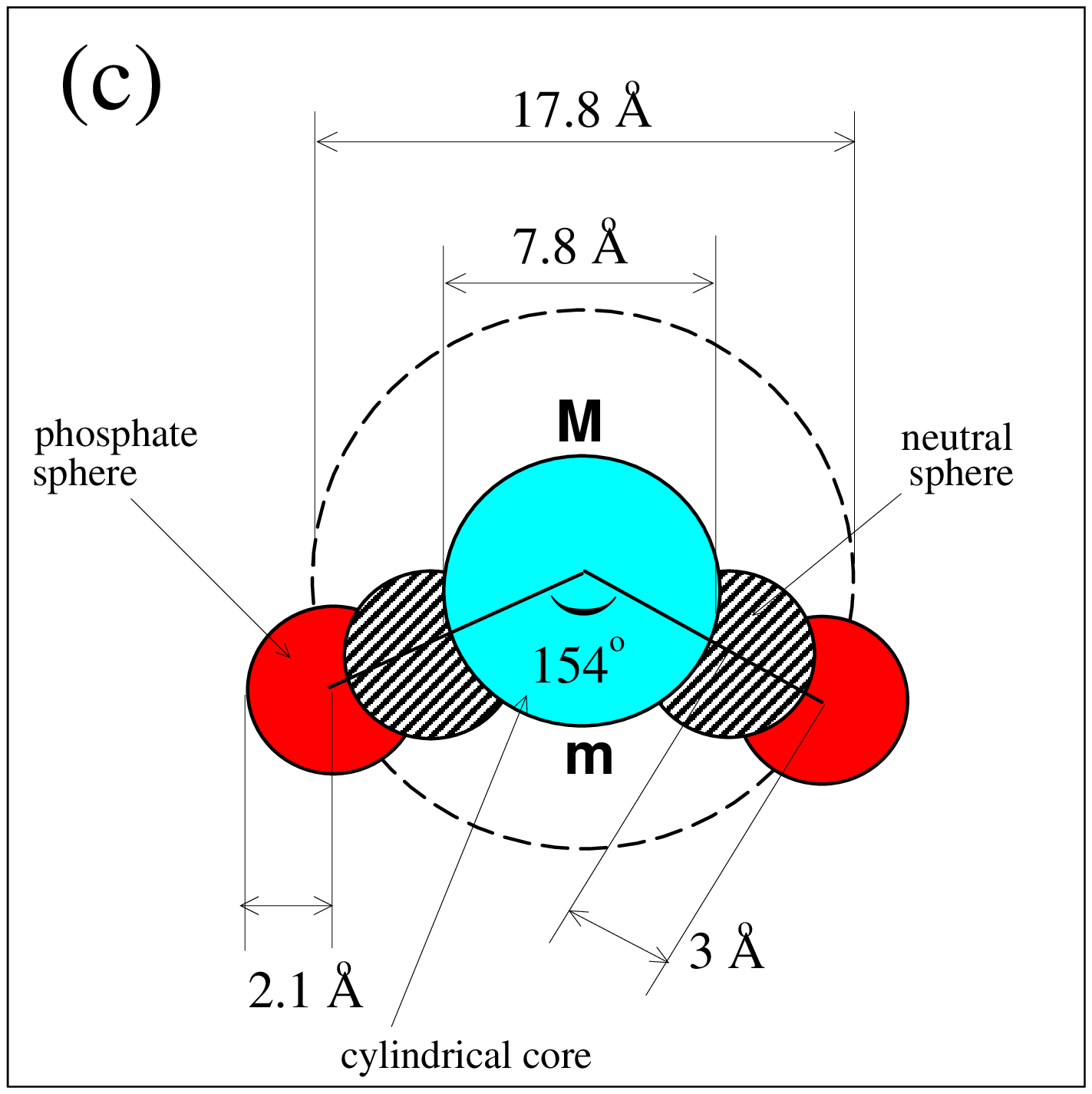} 
  \caption{Cross sections of different DNA models in the $xy$ plane. a) 
Cylinder model (CM), 
b) Extended cylinder model (ECM), c)
    Grooved, or Montoro-Abascal like model (MAM). Phosphate charges are
    shown as dark spheres. The DNA cylindrical core is colored in gray,
    the hatched areas correspond to neutral hard spheres. The
    inscribed letters "M" and "m" denote the major and minor grooves
    correspondingly. }
 \label{fig1}
\end{figure}

Three different models for the DNA shape will be studied here:
i) a simple  cylinder model (CM), ii) an  extended cylinder model (ECM) 
with a grooved structure and
iii) the Montoro-Abascal model (MAM) \cite{montoro1995}. The cross 
sections of these DNA
models in the $xy$ plane that is perpendicular to the DNA long axis 
($z$-axis)
and hits two phosphates on different strands are
sketched in Figure \ref{fig1} and will be discussed
subsequently:
in all three models the phosphate charges are discretely placed at
certain positions coincident with those of the phosphorus atoms in
crystalline DNA.

i) Cylindrical model (CM), see Figure \ref{fig1}a: This model was used 
by Kornyshev and Leikin 
\cite{kornyshevandleikin} and by two of us in another study 
\cite{ourfirstDNApaper}. 
In the CM, the cylindrical DNA core possesses
a diameter of $D$=20\AA. Two strings of point-like and monovalent 
phosphate
charges of size $d_p=0.4$\AA \, have cylindrical coordinates
($\rho_i^s,\phi_i^s,z_i^s$) relevant to
the phosphate sites of the B-form of DNA:

$\rho_i^s=D/2$=10\AA, \,\, $\phi_i^s = \phi_0^s + i\times 36^0$,
\,\,$z_i^s= z_0^s + i\times$ 3.4\AA. \\
Here $s$=1,2 specifies the nucleic acid strand, $i$=0,...,9 describes a
full DNA turn and the $z$-axis is the long DNA axis. Furthermore,
$\phi_0^1=0^0$, $z_0^1=0$ for the first strand ($s=1$) and  
 $\phi_0^2=144^0$, $z_0^2=0$ for the second strand ($s=2$). 

ii) Extended cylinder model (ECM), see Figure \ref{fig1}b: As designed by
Lyubartsev et. al. \cite{lyubar}, both  the helical DNA grooves and the 
discrete charge
localization on the DNA surface are incorporated. In the ECM, the DNA 
molecule has a hard cylindrical
core of diameter $D$=17.8\AA, \,\, which is slightly smaller  than in the 
CM. 
The phosphate charges are swollen up to hard sphere of diameter
 $d_p$=4.2\AA \, explicitly forming grooves. Other DNA parameters are 
similar to the CM.

iii) Montoro-Abascal model (MAM), see Figure \ref{fig1}c: This more
elaborate model was first 
introduced in Ref.\ \cite{montoro1995} where the grooved structure of
DNA is increased by adding another neutral sphere between the
cylindrical core and the charged 
phosphate sphere. In detail, the inner DNA cylindrical core of $D$=7.8\AA 
\,\, is overwinded by two strings of overlapping spheres. The outer string
of monovalently charged phosphate spheres are centered at a radial
coordinate of 8.9\AA. The radial position of inner string of neutral
spheres is 5.9\AA. Both spheres have the same $\phi$ and $z$
coordinates and diameter $d_p$=4.2\AA \,\, to incorporate a grooved 
geometry 
for the DNA molecule. Clearly, such a design of an overlapped
cylinder and two spheres creates a more grooved DNA profile with a deeper
cavity in the center of the minor groove. For other details of the MAM
and its reliability, we reference the reader to the original papers
\cite{montoro1995,montoro1998,montoro2001}.

\section{Simulation technique and system parameters}
In our simulation set-up, the 
B-DNA molecule is located in the center of a cubic simulation box. 
The cylinder axis is parallel to the $z$-axis and crosses the $xy$
plane at position ${\vec R}$ ($L/2,L/2,0$).
 The size $L$ of the simulation box was chosen to be
$L=102$\AA,\, corresponding to three 
full turns of B-DNA with a pitch length $P$=34\AA \,\, and with $N_p=3
\times 20=60$ phosphate groups along the DNA \cite{montoro1998}. There
is a small shift in the $z$-coordinate of two discrete phosphate
charges belonging to two different helices of $\Delta
z$=0.78\AA.

Periodic boundary conditions are applied in all three directions,
hence the DNA replicas in the $z$-direction produce an infinitely long
DNA molecule, and an infinite array of DNA replicas in neighboring
cells is simulated.
 The phosphate spheres are monovalent, i.e.\ 
their charge $q_p<0$ corresponds to one elementary charge $\vert e \vert$,
$q_p=-\vert e \vert$, and they have an effective diameter $d_p$ which is a
variable parameter in our different shape models. In addition to the DNA 
phosphates,
the system contains $N_c$ counterions 
with charge $q_c$, and $N_s\equiv N_+ \equiv N_-$ pairs of salt ions of
concentration $C_s=N_s/V^{'}$ with charges 
$q_+$ and $q_-$. Here $V^{'}$ is the free volume in the simulation box
accessible for these small ions, where the excluded volume of the DNA
molecule has been subtracted. The counterion number $N_c$
in the simulation box is fixed by the charge of 
the DNA molecule due to the constraint
of global charge neutrality, $N_c q_c=60 \vert q_p \vert $. For 
simplicity, we shall always deal
with a symmetric salt case, $\vert q_+ \vert = \vert q_-\vert$. 
All small ions are modelled as hard spheres of (hydrated ion) diameter 
$d_c$. For
most of our simulations, $d_c$=3\AA, but we have obtained data for larger
ion sizes $d_c$=6\AA \, and $d_c$=8\AA \, as well.
The whole system is held at room temperature $T=298K$. The dielectric
constant $\epsilon=78$ of the solvent is assumed to be uniform throughout
the suspension (same value inside the DNA molecule and in the suspending 
medium), which avoids electrostatic images.

\begin{figure}
 \vspace{10cm}
 \hspace{4cm}
    \epsfxsize=10cm
    \epsfysize=7cm 
  ~\hfill\epsfbox{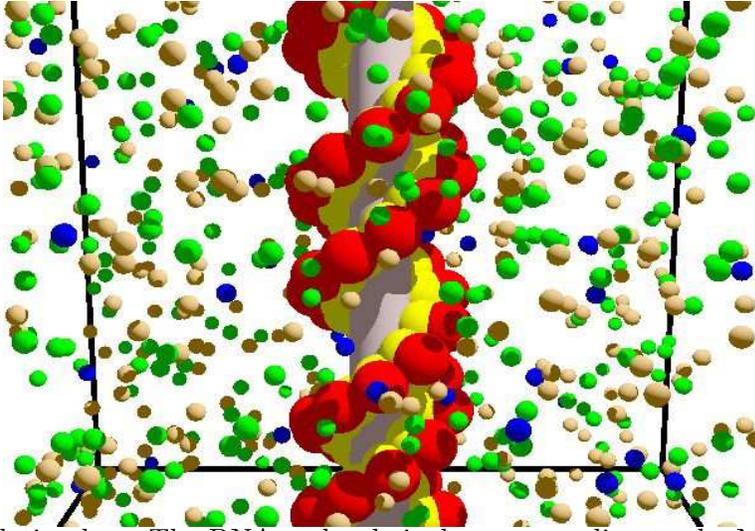}\hfill~ 
  \caption{Snapshot of the simulation box. The DNA molecule is drawn
    according to the MAM. Black spheres on the DNA strands represent
    the 
    phosphate charges. Internal grey spheres between the phosphates and
    the DNA cylindrical core are neutral. Positive (negative)
    salt ions spreaded across the simulation volume are shown as open 
(hatched)
    spheres.}
 \label{snapshot}
\end{figure}

The interactions between the mobile ions and the phosphate charges 
are described within the framework
of the primitive model as a combination of excluded volume and Coulomb
interactions reduced by the dielectric constant $\epsilon$ of the solvent.
The corresponding pair interaction potential between the different 
charged hard spheres is
\begin{equation}
V_{ij} (r) =\cases {\infty &for $ r \leq (d_i+d_j)/2 $\cr
  {{q_i q_j e^2} \over {\epsilon r}} &for $ r > (d_i+d_j)/2 $\cr},
\label{1cLMH}
\end{equation}
where $r$ is the inter-particle separation and $i,j$ are indices denoting
the different particle species. Possible values for $i$ and $j$ are
$c$ (for counterions), $+,-$ (for positively and negatively charged
salt ions), and $p$ (for phosphate groups).
In addition, there is an interaction potential $V^{0}_{i}$ between the DNA 
hard cylinder and the free ions
$i= c,+,-$ which is of simple excluded volume form such that these ions
cannot penetrate into the cylinder. Similar excluded volume potential
exist for the inner neutral DNA spheres in the elaborated MAM. 
Finally, the ionic strength $I$ and the Debye screening length $\lambda_D$ of
the solution are defined as $I=\frac {1}{2}(q_c^2 N_c/V^{'} + 
\sum_{j=+,-}q_j^2 C_s)$ and
$\lambda_D=\sqrt{{\epsilon k_B T}\over{4 \pi I}}$.  
In order to compute the statistical averages over the
mobile microions, we
 have performed conventional $NVT$ molecular dynamics simulations,
where the long-range electrostatic forces were treated
according to the Lekner procedure \cite{ourfirstDNApaper,lekner}.
A typical simulation snapshot of the system is given in
Figure \ref{snapshot}. 

Our major goal is to calculate the mobile  ion number densities
$\rho_j(\vec r)$ $(j=c,+,-)$ around the DNA molecule. They are defined as 
a statistical average,
\begin{equation}
\rho_j({\vec r})= \langle \sum_{i=1}^{N_j} \delta ( {\vec r} - \vec
 r_i^{j} )\rangle.
\label{denc}
\end{equation}
Here $\{ {\vec r}_i^{j}\}$ denote the positions of the $i$th particle of 
species $j$.
 The canonical
average $<...>$ over an $\{ {\vec r}_i^{j} \}$-dependent
quantity $\cal A$ is defined via the classical trace
\begin{eqnarray}
\langle {\cal A} \rangle = &&{1\over
 {\cal Z}}\Bigl\{ \prod_{k=1}^{N_c}\int d^3 r_k^c \Bigl\}
\Bigl\{ \prod_{m=1}^{N_+}\int d^3 r_m^+ \Bigl\}
\Bigl\{ \prod_{n=1}^{N_-}\int d^3 r_n^- \Bigl\}
 \nonumber \\
&& {\cal A} \exp
 \lgroup -\beta\sum_{i=c,+,-} [ V^{0}_{i} + \sum_{j=c,p,+,-} U_{ij} ]
 \rgroup.
\label{33}
\end{eqnarray}
Here $\beta= 1/k_BT$ is the inverse thermal energy ($k_B$ denoting
Boltzmann's constant) and 
\begin{eqnarray}
&&U_{ij} =(1-{1 \over 2}\delta_{ij})\sum_{l=1}^{N_i} \sum_{k=1}^{N_j} 
V_{ij}( \mid {\vec r}_l^i - {\vec r}_k^j \mid ), 
\label{9999}
\end{eqnarray}
is the total potential energy of the counter- and salt ions provided the
phosphate groups are at positions $\{ {\vec r}_n^p\}$ ($n=1,...,N_p$). 
Note that the periodically repeated particles are incorporated implicitly
in the interaction energy. Finally the
prefactor $1/{\cal Z}$ in Eq.(\ref{33}) ensures a correct normalization,
 $<1>=1$.

\begin{figure}
 \hspace{4cm}
    \epsfxsize=10cm
    \epsfysize=10cm
 ~\hfill\epsfbox{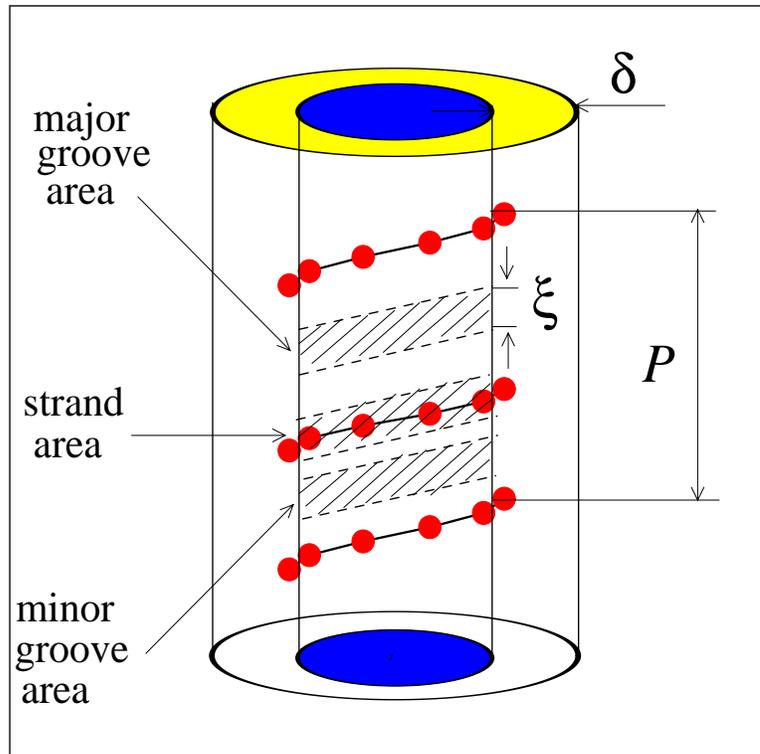}
  \caption{A schematic picture to explain the procedure of
     ion density calculations along one pitch length (P) of a DNA
     molecule. The filled circles connected with full line are phosphate
groups. The shaded areas correspond to a  path along the major 
     groove, minor groove  and one  of the phosphate strands. 
The path height is $\xi=3.4\AA$ \,\, and width is  $\delta$=5\AA. }
 \label{condensshell}
\end{figure}

 In computer simulations and different theoretical
approaches the distance below which the ions are considered to
be condensed is usually assumed to be in the range of one or two water 
molecule
diameters. Thus the width of the condensation shell near the DNA surface
is around 7\AA \,\, in Manning
theory \cite{manning1}, whereas a value of 5\AA \,\, was invoked in other
papers \cite{feig1999,lyubar,andreasson1993,braunlin1987}. These values 
are
larger than the thickness of the Stern
layer $d_l=A/4\pi \lambda_B$=2\AA \,\, including ions bound to the 
molecular
surface. Here $\lambda_B=e^2/\epsilon k_B T$ is the Bjerrum length and $A$ 
is an
average area per elementary charge on the molecular surface.
In our study we follow the latter criterion and treat the 
ions as condensed if the {\it surface-to-surface distance} between the ion 
sphere and
the DNA hard surface is not larger than 2\AA. 
In other words, we are interested in the population in special areas
of the DNA surface by small ions rather than the actual ion condensation
on the DNA surface \cite{Holm2000}.
In order to resolve the adsorbed ions along the strands and in the major
and minor grooves, we integrate the ion density fields as given by Eqn.(2)
over  a small volume close to the DNA surface. This volume is bounded
by the parallel surface to the DNA
surface at distance $\delta=$2\AA$+d_c/2$ and has a height $\xi$ in $z$-direction.
The volume follows the helical symmetry of the DNA molecule.
A schematic view of this condensation shell and the definition
of the groove and strand adsorption paths around the DNA are
given in Figure \ref{condensshell}.  We have separately counted cations, 
that
 comprise of counterions and positively charged salt ions, and anions
 (coions). The angular-resolved cation and anion density
profiles along the phosphate strands and the minor and major grooves 
represent the main results of this paper. We call a plot of the densities versus
polar angle a ``panoramic'' view of the density profiles and 
shall present  exhaustive data for these quantities for different 
parameters 
in the next section. Additionally  we define the charge densities of
adsorbed ions by 
 $\rho^{(+)} = \sum_{j=c,+} q_j  \rho_c^{(j)}$ and $\rho^{(-)} = q_-
 \rho_c^{(-)}$, where $\rho_c^{(j)}$  are the number
 densities of the adsorbed ions of species $j \in \{c,+,-\}$.

Our molecular dynamics simulations cover a broad range of salt
concentrations from
0.1Mol/l to 1.61 Mol/l, where the latter corresponds to  2000 salt
ions of both charges in the simulation box. Details of the simulated 
states are 
summarized  in Table I. During the simulation we checked
that there was a continious exchange of the adsorbed and free ions,
which demonstrates that our systems are in equilibrium.

\section{Results}
\subsection{Monovalent ions}

\begin{figure}
 \hspace{-2cm}
 ~\hfill\epsfbox{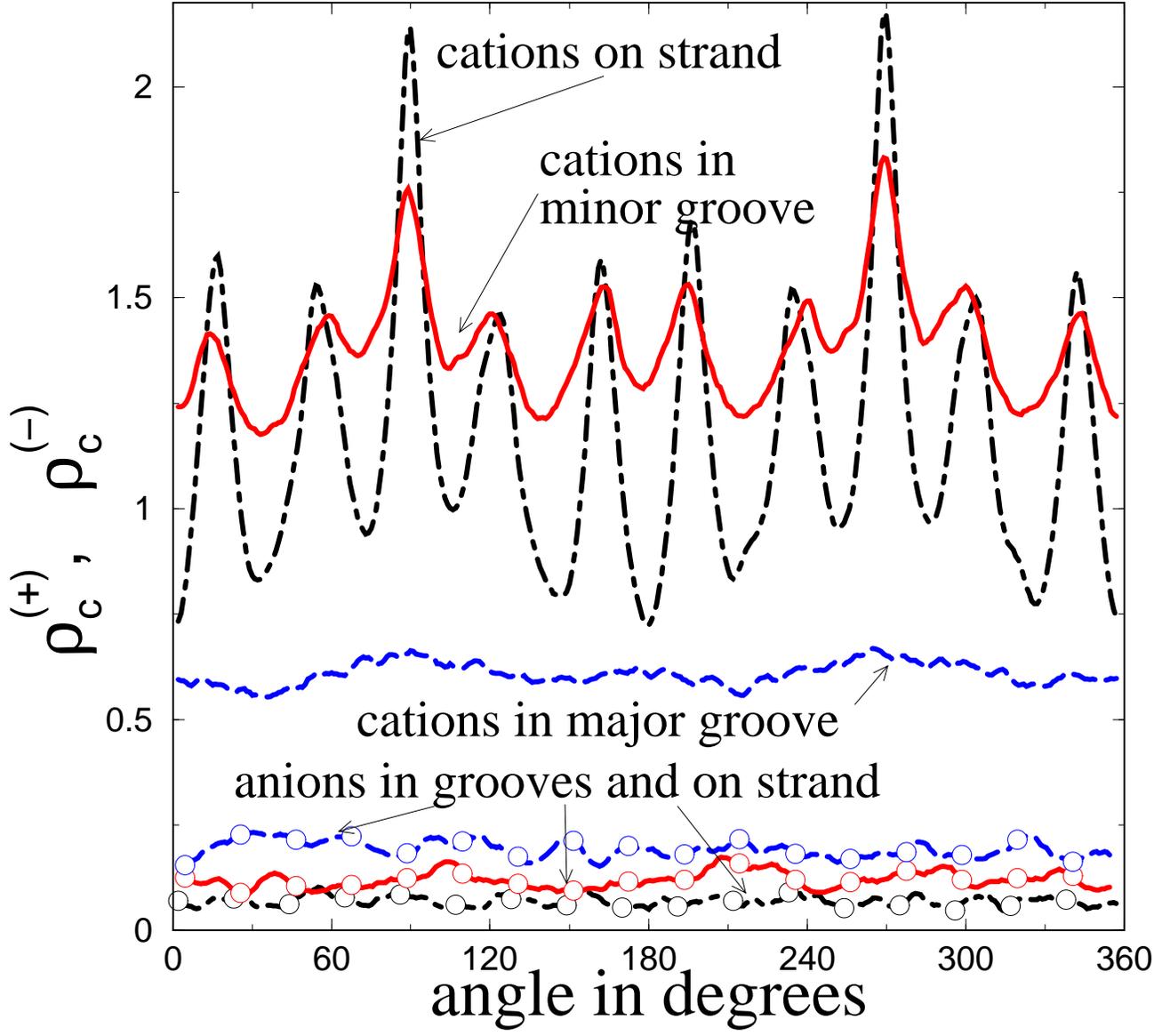} \hfill~ 
  \caption{Panoramic view of the condensed small ion densities near the
    DNA surface for $q_c=1$, $q_s=1$ (Set 1), $C_s=0.1$Mol/l and the
    ECM. The density unit 
is 0.02 Mol/l. Dot-dashed lines: distribution on
the phosphate strands, full lines: distribution in the minor groove, 
dashed lines:
distribution in the major groove. Lines without or with symbols
correspond to cation $\rho_c^{(+)}$ or anion $\rho_c^{(-)}$
densities. The value of the cation distribution is much
larger than the anion distribution (where the latter is enhanced by a factor
of 10).}
 \label{fig4}
\end{figure}

\begin{figure}
    \epsfxsize=15cm
    \epsfysize=15cm 
 ~\hfill\epsfbox{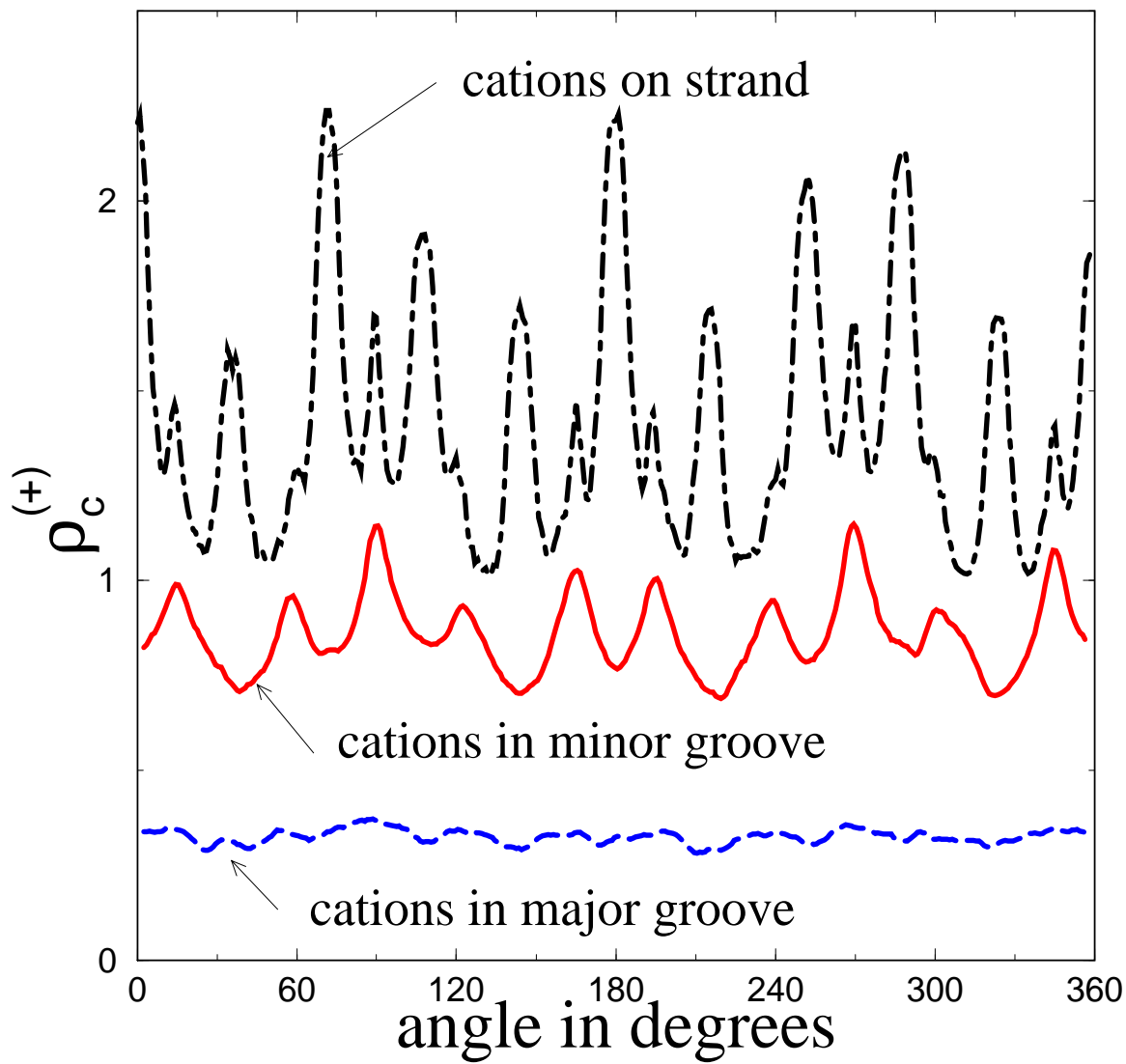}\hfill~
  \caption{Panoramic view of the condensed small ion densities near the
    DNA surface for the cylinder model (CM). The parameters are the
    same as in Figure \ref{fig4}.}
 \label{fig5}
\end{figure}

\begin{figure}
    \epsfxsize=15cm 
    \epsfysize=15cm
 ~\hfill\epsfbox{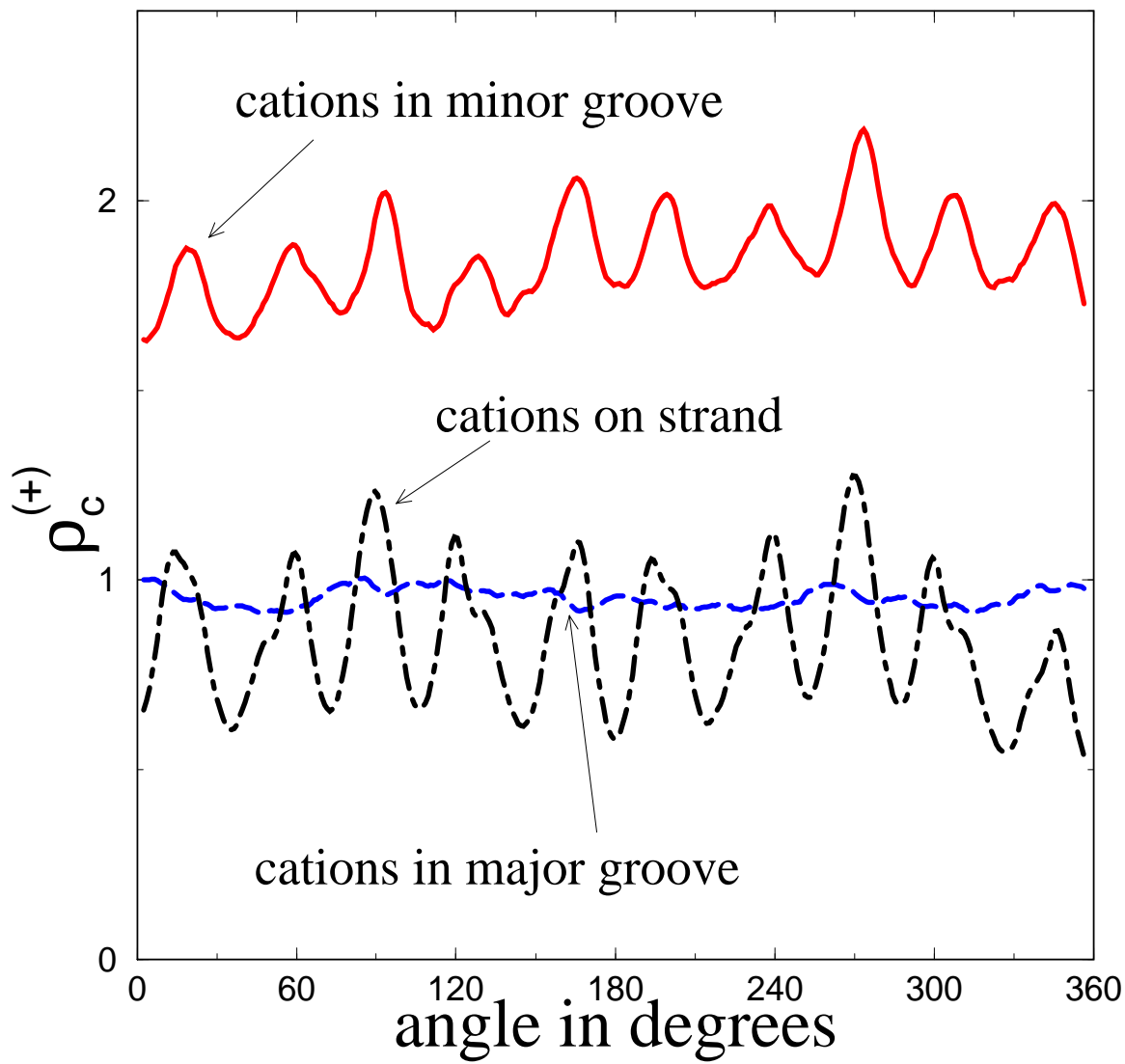}\hfill~ 
 \caption{Panoramic view of the condensed small ion densities near the
    DNA surface for the Montoro-Abascal model (MAM). The parameters are the
    same as in Figure \ref{fig4}.}
 \label{fig6}
\end{figure}

Let us first discuss monovalent ions. For the extended cylindrical model 
(ECM),
a  panoramic angular cat- and anion distribution over $0<\phi<2\pi$
is plotted in Figure \ref{fig4}  for
the system parameters $q_c=1$, $q_s=1$ (Set 1 of Table I) and  $C_s=0.1$Mol/l.  
The cations cluster in front of the charged phosphates, almost in
a site-binding like manner
\cite{korolev1999,montoro1995,montoro1998,montoro2001}, showing a
strong structuring, while
the minor-groove cation density is less structured.
The smallest cation density is in the major groove. The anion
densities in the DNA grooves and on the phosphate strands are 
 considerably smaller and not structured at all. The major groove
 population of anions is higher than that in the minor groove and on
 the strands. 
The same quantities are shown for the cylinder model (CM) and the 
Montoro-Abascal 
model (MAM) in Figures \ref{fig5} and \ref{fig6} respectively.
As compared to the ECM, an increase of condensed cations along the 
phosphate strands
at the expense of their accumulation in the minor and major grooves is
clearly visible for the CM. In the most realistic MAM, however, more 
cations bind and locate in the DNA grooves \cite{howerton,mcf}  at the 
expense of their
accumulations on strands. Such an ion relocation from the strands into
the grooves entails an entropy gain for salt ions. 
 In the MAM the population of condensed cations along the strands is 
less by a factor of two as compared to the structured minor groove 
density. 
 Our conclusion arising from Figures   
  \ref{fig4}, \ref{fig5}, and \ref{fig6} is twofold:
 first, on a technical level, the inclusion
of a grooved shape in the excluded volume of the DNA molecule is crucial 
for
ion adsorption. It completely changes the charge and structure of the 
adsorption pattern.
Second, taking the MAM as the most realistic description of the DNA shape, 
we
can conclude that adsorbed cations exhibit a pronounced spatial structure 
along the minor
groove. This strong structuring of cations
in the minor groove will induce also a structuring of the water molecules 
in the 
minor groove and might therefore be related to the so-called ``spine of 
hydration in the minor groove''
which is attributed to a high water ordering there 
\cite{feig1998,halle1998,denisov2000,soler-lopez1999,mcconnell2000,schneider1998}. 
 In general, the spine of
hydration emerges due to the occasional intruding of counterions
to the particularly electro-negative regions in the
minor groove \cite{yang1997jacs} and is addressed in
Refs.\cite{Jayapura1989,montoro1998,korolev2001biopolymers,korolev2002biophys,mcconnell2000,Pack1993,rajasekaran1994,young1997}.
We think that the experimentally
measured hydration pattern can be a fingerprint of cation ordering in the 
minor
groove \cite{soler-lopez1999,shui1,shui2}.

\begin{figure}
    \epsfxsize=15cm
    \epsfysize=15cm 
 ~\hfill\epsfbox{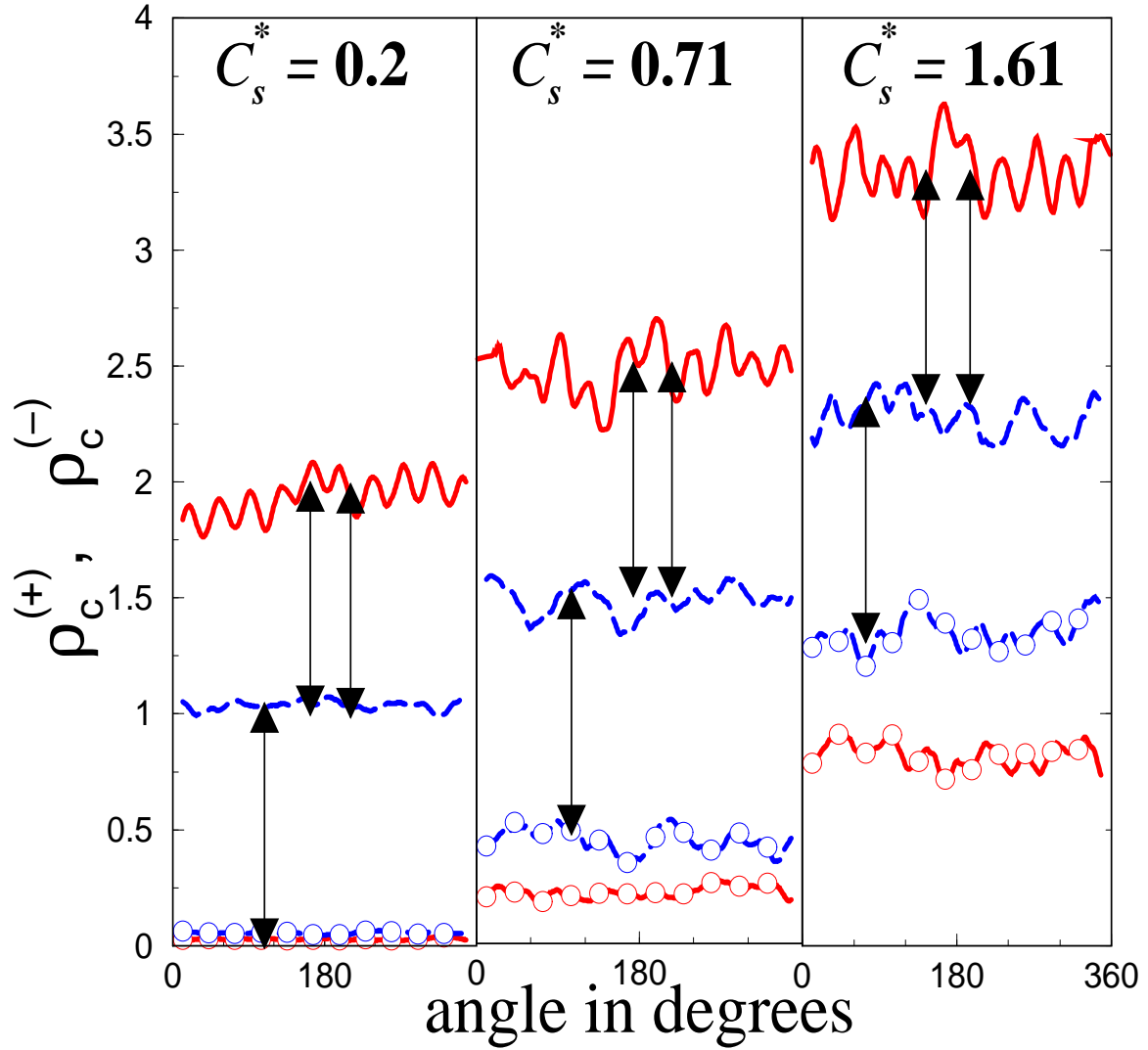}\hfill~
 \caption{Panoramic view of ion densities in DNA grooves for MAM and
   $q_c=1$, $q_s=1$ (Set 1). Three full
   panoramic views along the {\it x} axis correspond to three
   different salt densities; from left to right,
   $C_s^*=C_s/C_s^0=$0.2, 0.71, 1.61 with $C_s^0= 1$Mol/l.  Full
   line- charge distribution in the minor groove, dashed line- charge
   distribution in
 the major groove. Lines without or with symbols
correspond to cation $\rho_c^{(+)}$ or anion $\rho_c^{(-)}$ densities. The single arrow
   indicates the constancy of the major groove charge at different added
   salt densities. The constancy of the difference between the cationic
   charges of DNA grooves is shown as a double arrow.}
 \label{fig7}
\end{figure}

The dependence of the adsorption pattern
on the salt concentration is shown in Figure \ref{fig7}.
The added monovalent salt concentration  $C_s$ is increased from 0.2 Mol/l 
to 1.61 Mol/l for $q_c=1$, $q_s=1$ (Set 1).
The highest cationic
occupation is again in the minor groove but more anions condense
in the major groove as the salt density is increased \cite{montoro2001}.
The total major
groove charge, calculated as a difference between the cation (dashed line)
 and anion (dashed line with symbols) densities in the major groove, is 
almost independent of the amount of added salt, see the single arrows in Figure
\ref{fig7}. In other words, for any monovalent salt 
density 
the geometry of major groove and the electrostatic field of the two
adjacent phosphate strands regulate the cation and anion populations
and keep the major groove charge unchanged. 
On the other hand, the minor groove is positively charged as the bulk
salt density increases, see the difference between the full line and
the full line with symbols for different salt densities in Figure
\ref{fig7}. 
The visible cation structuring in the major groove at dense salt, see
the right side of Figure \ref{fig7}, is
consistent with the experimental evidence for recurring hydration patterns
in the major groove \cite{young1997,tippin,eisenstein1995}.
The other
observation is a constancy of the gap between the minor and the
major groove cationic
occupancies  shown in Figure \ref{fig7} by the double arrows for
different salt densities. Obviously in solutions, where the
DNA phosphate charges are
effectively screened out, it is the osmotic pressure of salt 
that pushes ions close to the DNA surface. We note that the constancy of
the accumulated charge in the major groove and the constancy of the
difference between cationic
populations of major and minor grooves do not appear in the ECM and CM. 

We finally remark that throughout all runs it was revealed that, a
distance 5\AA\,-7\AA\, away from the DNA surface a cylindrical
symmetry of
radial ion distribution is completely restored in accordance with
observations of
Refs. \cite{lyubar,montoro1998,gulbrand1989,jaya2}. Thus the effect of
discreteness of DNA charges on counterion concentration
profile is generally small and dwindles a few angstroms away from the DNA
surface \cite{conrad1988biopol,hochberg}.

\subsection{Multivalent counterions and monovalent salt}
\begin{figure}
    \epsfxsize=15cm
    \epsfysize=15cm 
 ~\hfill\epsfbox{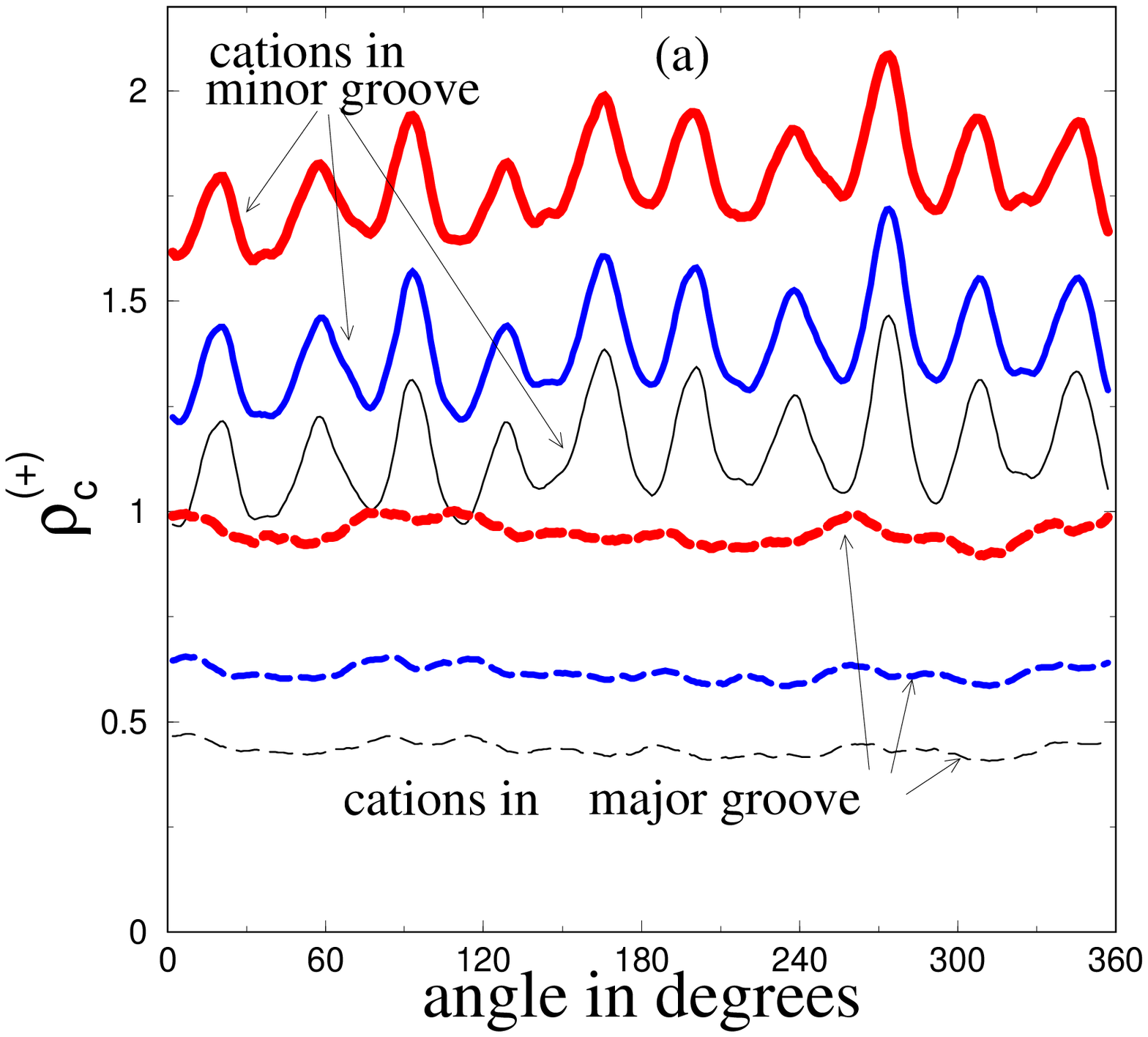} \hfill~ \newpage
    \epsfxsize=15cm
    \epsfysize=15cm 
 ~\hfill\epsfbox{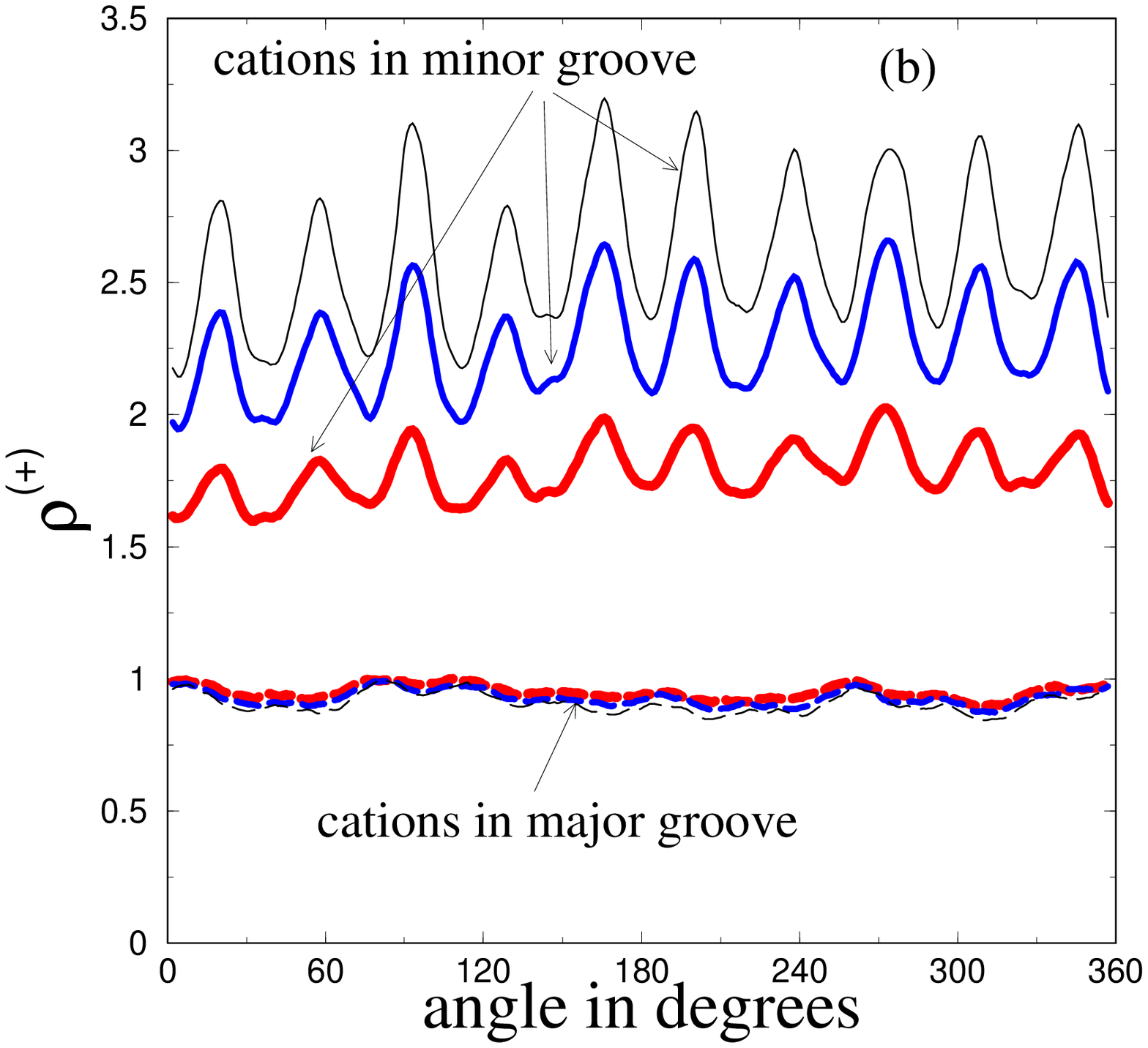}\hfill~
  \caption{Panoramic view of the cation number density $\rho_c^{(+)}$ (a) and 
charge density
    $\rho^{(+)}$ (b) for the MAM. Monovalent salt
    $C_s=$0.1 Mol/l and
    different counterion valencies.
 Full line- cations in the minor groove,
   dashed line- cations in the major groove.  
Thick lines- monovalent counterions (Set 1),
    medium sized line- divalent counterions (Set 4),
    thin line- trivalent counterions (Set 5).  
} 
 \label{fig9}
\end{figure}

We now consider the case of multivalent counterions 
and a monovalent salt, for which the Coulomb
correlations between the counterions and the DNA phosphates are strong.
We keep  the  salt
concentration fixed and increase the counterion
valency. This leads to a higher on-strand adsorption
of couterions which implies less condensation in the minor
and major groove, see Figures \ref{fig9}a.
Also for high counterion valencies, the cation adsorption on the minor groove
is higher than that on the major groove in accordance with Ref.\ \cite{stigter1998}.
The same trends appear also in the ECM and in the CM.

Furthermore, for increasing counterion valency, Figure \ref{fig9}b reveales that
the total adsorbed charge in the major groove is {\it almost constant} while
it is getting more positive in the minor groove.  The increase of
the adsorbed cations in the minor groove causes a visible spatial
structuring along the minor groove, see the surging
oscillations in Figures \ref{fig9}a and  \ref{fig9}b. Again, such
an ion structuring is perhaps connected to the experimentally observed
spine of hydration. Note that the number of adsorbed ions in the major groove 
drastically decreases in the CM and ECM for higher counterion valencies,
which reaffirms the crucial role of the modelling of the 
DNA shape.  

\subsection{Multivalent counterions and multivalent salt}
\begin{figure}
    \epsfxsize=15cm 
    \epsfysize=15cm 
 ~\hfill\epsfbox{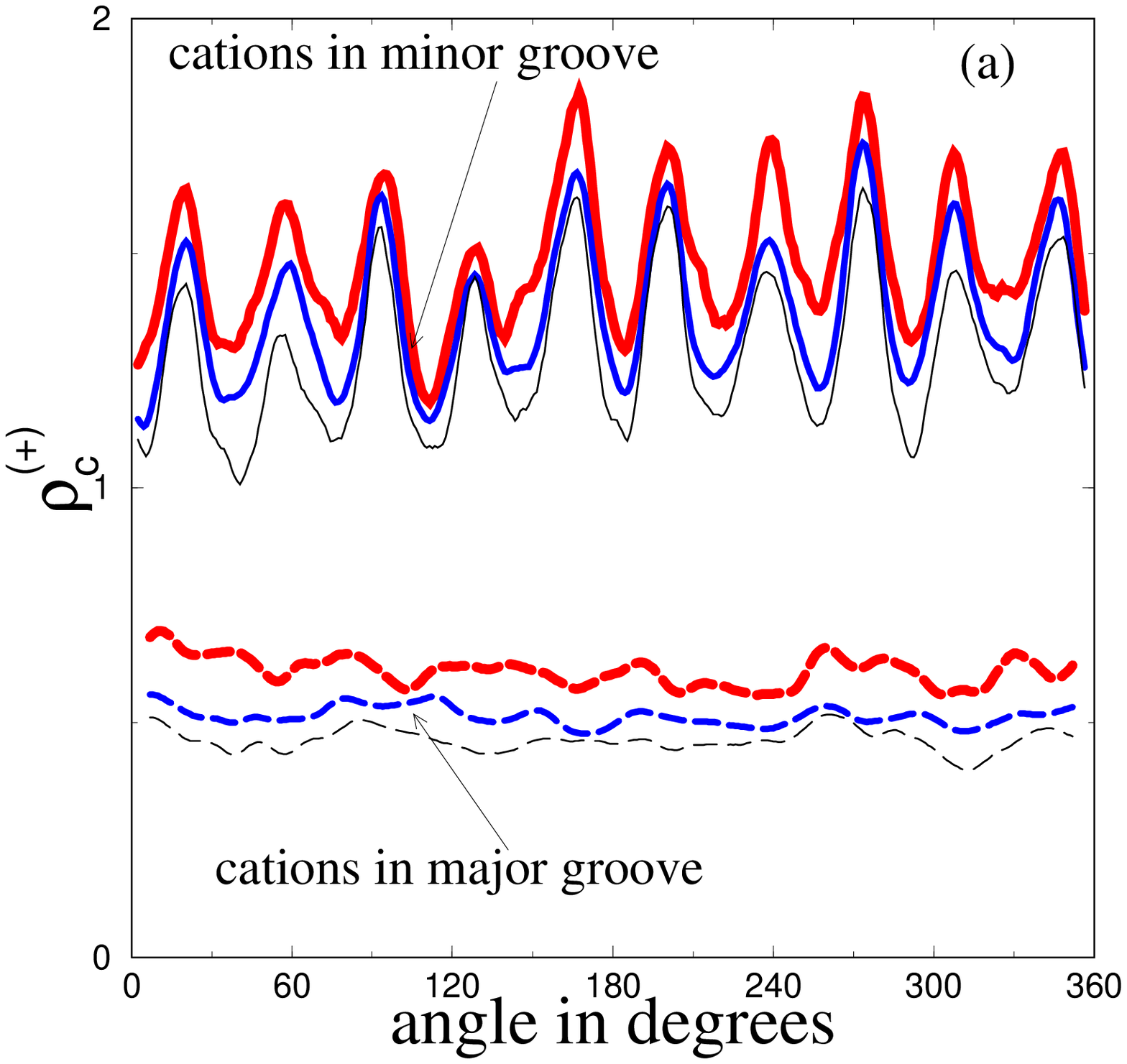} \hfill~ \newpage
    \epsfxsize=15cm
    \epsfysize=15cm 
 ~\hfill\epsfbox{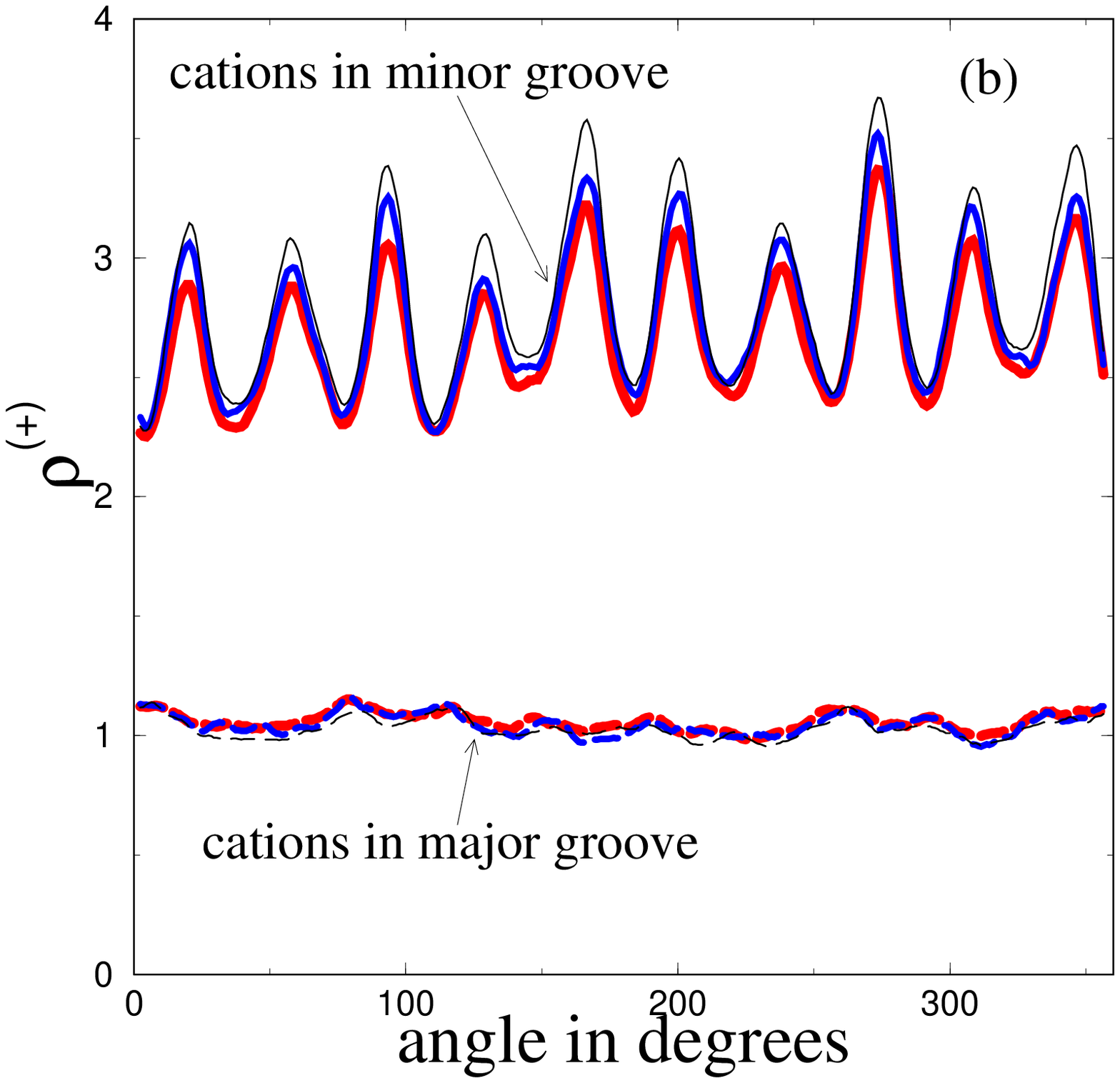}\hfill~
  \caption{Panoramic view of the cation number density $\rho_c^{(+)}$ (a)
    and charge density $\rho^{(+)}$ (b)  for the MAM. 
 Divalent salt $C_s=$0.2 Mol/l and
 different counterion valencies.  
Full lines- cations in the minor
    groove, dashed lines- cations in the major groove.
     Thick lines- monovalent counterions (Set 2),
    medium sized line- divalent counterions (Set 6),
    thin line- trivalent counterions (Set 3).
} 
 \label{fig10}
\end{figure}

For  fixed {\it divalent} salt concentration and increasing
counterion valency, similar results are obtained as shown in
Figure \ref{fig10}. The number of adsorbed cations decreases in the grooves,
see Figure \ref{fig10}a, the total charge in the major groove stays constant
(very close to its value in Figure \ref{fig9}b)
while it is rising slightly in the minor groove, see Figure \ref{fig10}b.
Hence we conclude that total major groove charge is independent of
valencies of both cations and anions.

\begin{figure}   
 \epsfxsize=15cm 
    \epsfysize=15cm 
 ~\hfill\epsfbox{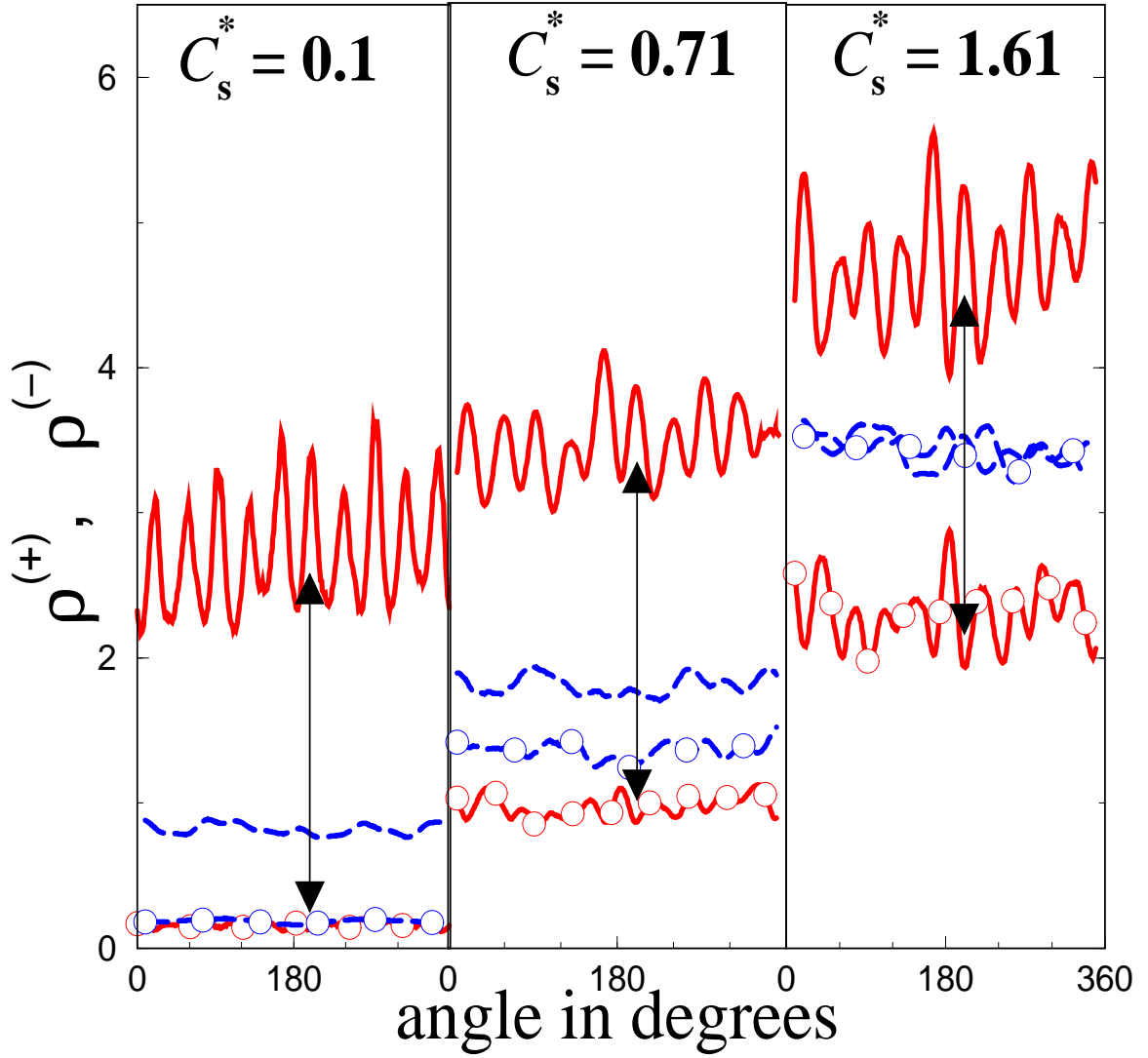}
   \caption{Panoramic view of ion charge densities in the DNA grooves
     for the MAM, trivalent 
     counterions and divalent salt (Set 3) and three different salt
     concentrations; from left to right, $C_s^*=C_s/C_s^0=$0.1, 0.71, 1.61
with $C_s^0= 1$Mol/l. 
 Full
   line- charge distribution in the minor groove, dashed line- charge
   distribution in
 the major groove. Lines without or with symbols
correspond to cation $\rho^{(+)}$ or anion $\rho^{(-)}$ charge densities.
The shrinking of the gap between the major groove cationic (dashed
line) and anionic (dashed line with symbols) charges, as more salt is
added, is the onset of the major groove neutralization. The minor groove
charge does not depend on the salt concentration, see arrows 
which indicate the total charge density in minor groove (the gap between the minor
     groove cation (full line) and anion (full lines with symbols) charges).}
 \label{fig12}
\end{figure}

\begin{figure}
 \hspace{4cm}
    \epsfxsize=15cm 
    \epsfysize=15cm 
 ~\hfill\epsfbox{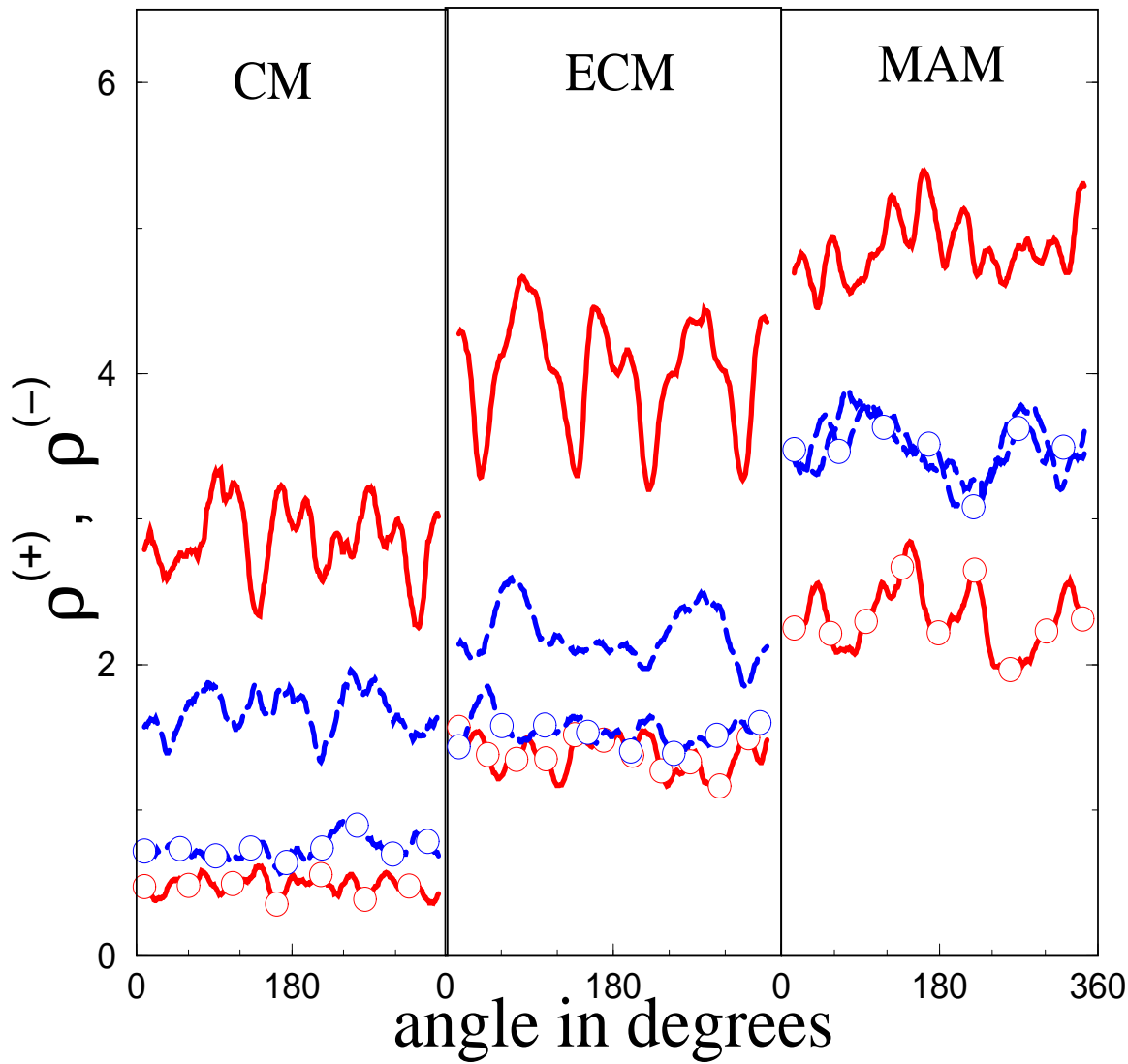}
  \caption{Panoramic view of ion charge densities in the DNA grooves, 
monovalent counterions and divalent salt (Set 2) and $C_s=$1.61 Mol/l. 
   Different DNA models, from left to right: CM, ECM, MAM.
 Full line- charge distribution in the minor groove, dashed line- charge
   distribution in
 the major groove. Lines without or with symbols
correspond to cationic $\rho^{(+)}$ or anionic $\rho^{(-)}$ charge densities.
Note that the major groove
   neutralization, described by a coincidence of the major groove
   cationic (dashed line) and major groove anionic (dashed line with
   symbols) densities, appears only in the MAM.}
 \label{fig11}
\end{figure}

We now increase gradually the amount of added divalent salt. 
The in-groove ion distributions are shown in Figure \ref{fig12}
for three different salt concentrations.
There are two remarkable effects for increasing salt concentration: first
the total charge of ions adsorbed in the {\it major} groove is
approaching zero, i.e.\ {\it the major groove is neutralized} for high salt concentration. Second,
the total charge in the {\it minor} groove is pretty robust
against an increase in salt concentration, see the length of the 
arrows in Figure \ref{fig12}.
The effect of major groove neutralization are lost if a less realistic
DNA shape is used as shown in Figure \ref{fig11} for the CM and the ECM.

\begin{figure}   
 \epsfxsize=15cm 
    \epsfysize=15cm
 ~\hfill\epsfbox{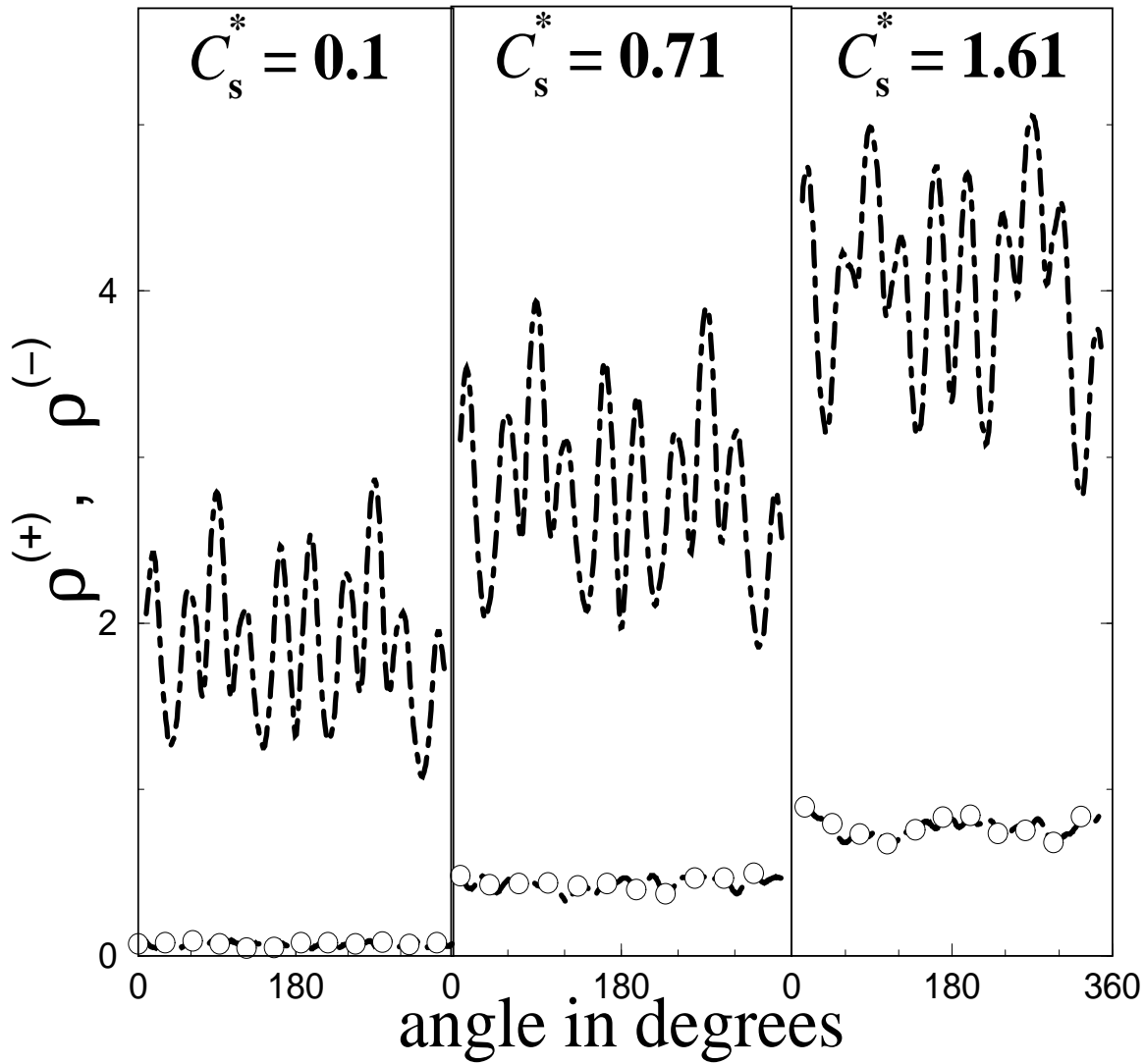}
   \caption{Panoramic view of ion charge densities on the phosphate strands for 
the MAM, trivalent counterions and divalent salt (Set 3) and three
different salt concentrations; from left to right, 
 $C_s^*=C_s/C_s^0=$0.1, 0.71, 1.61 with $C_s^0= 1$Mol/l. 
The total ionic charge on strand, defined as a difference between the
cationic charge $\rho^{(+)}$ (dot-dashed line) and anionic charge
$\rho^{(-)}$ (dot-dashed
line with symbols) increases as more salt is
 added to solution.}
 \label{fig13}
\end{figure}

In Figure \ref{fig13}, the ion densities adsorbed on the strands are shown for
trivalent counterions as a function of divalent salt concentration.
The total charge adsorbed along the strands increases with added salt
concentration. Togther with the constancy of the minor groove and the
neutralization of the major groove this produces an overcharging effect of the DNA
which is discussed in the next subsection.

\subsection{Overcharging effect}
Charge inversion (also known as an overcharging,
overneutralization or charge reversal)  is possible for
a variety of macroions, ranging 
from the charged surface, charged lipid membranes to colloids, DNA and 
actin. 
 It is believed that for this effect to occur, the cations have to be
multivalent to enhance the nonlinear effects, such as  Coulomb
correlations.
Thus, in the presence of multivalent ions, the ionic cloud may not only
compensate the polyion charge, but even exceed it, resulting in
opposite values of the electrostatic potential at some distances. 
Overcharging
has been observed in Monte-Carlo simulation
\cite{vlachy,torrie1980,degreve,jonsson2001}, HNC calculations
\cite{tovar1985jcp,greberg1998and2001,kjellander} and modified PB theories
with  nonlinear correlations included 
\cite{park1999,joanny1999,leermakers2003}.
In detail, we consider a "physical" overcharging 
\cite{montoro2001,deserno2003},
when the sign of total charge of the complex of macroion and small ions,
which are localized in a thin shell
around the macroion, is opposite to the sign of bare macroion charge.
 Our definition of  overcharging  
 differs therefore from the counterion 
 or so-called $Z$-ion induced "structural" overcharging
\cite{messinaovercharge,patra2002,shklovskii2000,shklovskii2001,talingting,levin1999} 
which does not account for the adsorbed salt ions and 
disappears as more salt is added to solution \cite{allahyarovsolventeffect}.

\begin{figure}
   \epsfxsize=15cm 
   \epsfysize=15cm 
 ~\hfill\epsfbox{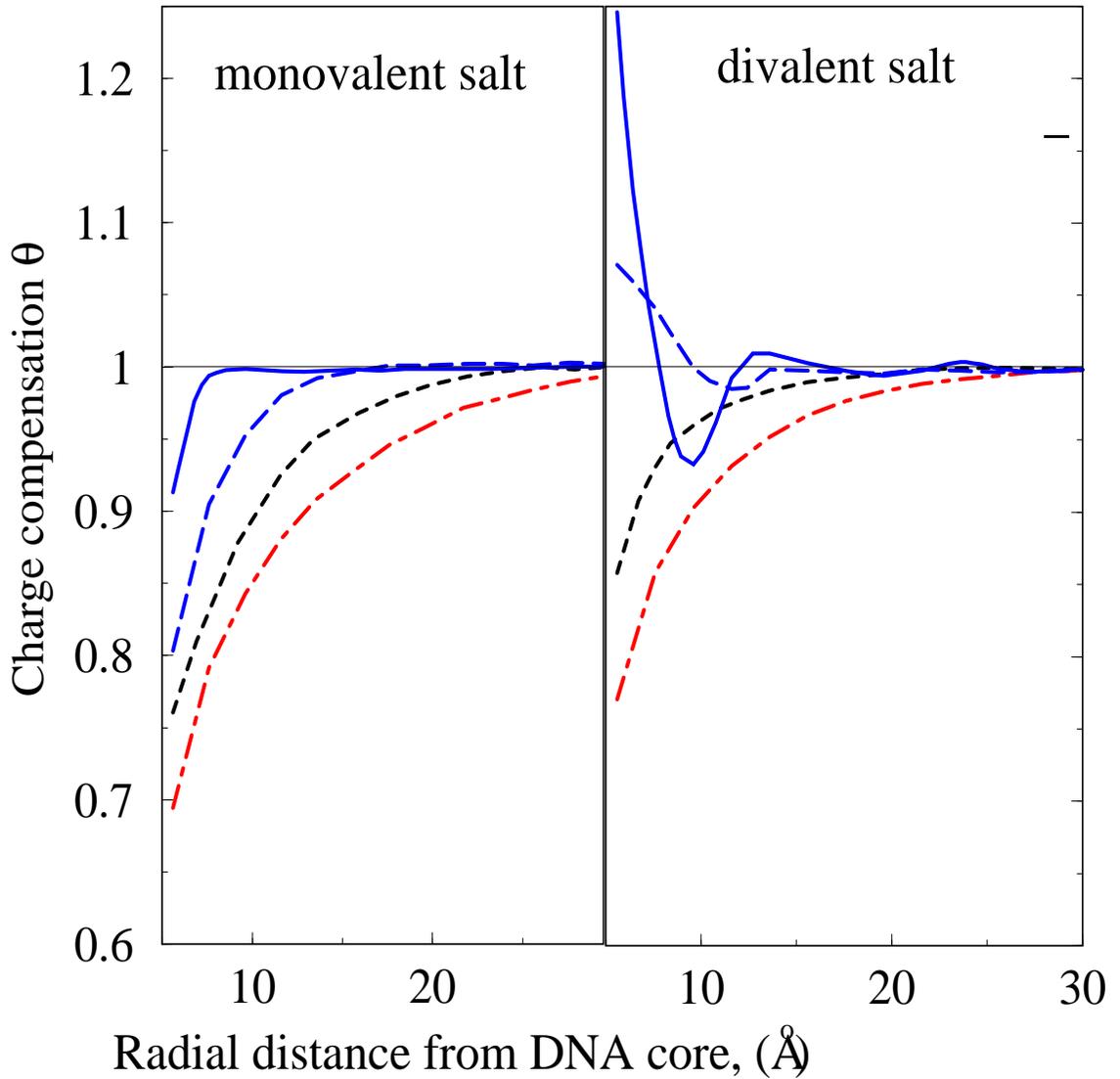}
   \caption{Charge compensation parameter $\theta$ versus distance
     from the DNA core surface for the MAM and different salt
 densities. The salt concentration $C_s$ is increased from bottom to top;
 0.1 M/l (dot-dashed line), 0.2 M/l (dashed line), 0.71 M/l (long
 dashed line) and 1.61 M/l (full line).
 (a) trivalent counterions and monovalent salt (Set 5). 
 (b) monovalent counterions and divalent salt  (Set 2).
}
 \label{fig14}
\end{figure}

 During the simulations the charge compensation parameter of DNA
 phosphate charges, defined as 
\begin{equation}
\theta(r) = \left(q_c \rho_c(r) + q_+ \rho_+(r) + q_- \rho_-(r)
  \right) / N_p |q_p| \, ,
\end{equation}
 was calculated. Here $N_p=60$ is the number of phosphate charges in
 the simulation box. The parameter $\theta(r)$ accounts for the integrated total 
charge
 at distance $r$ away from the DNA surface and has the following physical
 meaning.
For  $\theta(r)<1$, the DNA molecule is seen as a negatively charged
 rod at distance $r$ from its surface. Otherwise, if 
$\theta(r)>1$, the effective DNA charge at distance $r$ from its
 surface is positive. 
Data for $\theta$ are plotted in Figure \ref{fig14} for the MAM and 
different salt densities. The denser the salt, the stronger
the DNA screening. A qualitatively similar picture to
Figure \ref{fig14} appears for the ECM and CM and different counterion
valencies. There is no DNA overcharging in a solution of monovalent salt 
and multivalent counterions, see the left-hand side of Figure \ref{fig14} .
 For divalent salt,which is shown in the right-hand side of Figure
 \ref{fig14}, and low salt densities the compensation
parameter is  monotonic, resembling the monovalent salt case. 
However for dense salt $\theta > 1$ in the DNA
 vicinity. At the highest salt density involved in our simulations,
 $C_s$=1.61 Mol/l, there are  
 even several subsequent overcharging layers:  within the layer closest to
DNA surface the effective charge is positive, then within the
second layer the effective DNA charge is
negative, and finally within the third layer the effective charge
 becomes again positive, see the full line in the right-hand side of Figure \ref{fig14}. 
We mention that similar overcharging pictures are obtained
 for the divalent and trivalent counterions in solution with a divalent 
salt. Thus,
 summarizing the results of Figure \ref{fig14} we
 arrive at the conclusion that it is rather the multivalency of salt ions,
 than the multivalency of counterions which governs DNA
 overcharging. Our simulations also show that an
overcharging with divalent salt in the CM pops up only for multivalent
counterions.

\subsection{Varying the ion radius}
 The ion size has the meaning of a hydrated ion diameter and is an
adjustable parameter of the model that includes effects of the
molecular nature of the solvent in an averaged sense. 
The monovalent counterions are less solvated than divalent cations,
the latter condense via their solvent
ligands to H-bonds on a DNA surface 
\cite{chiu2000jmb,hydrogenbonds,heath1992}. Since we omit
the
ion chemisorption and do not account for specific ion
effects as exemplified by the Hofmeister effect
\cite{ninham1997langmuir,tardieu2002,piazza2000,philippe2002},
 only an electrostatic interaction with the phosphate
backbone is taken into account. A small change of the ion size, just in 
the range of the
typical values of the hydrated ion diameters, may cause a transition
to an attractive DNA-DNA interaction with a
spontaneous assembly of DNA system into an ordered
phase \cite{lyubartsev1995}. Therefore the hydrodynamic ion size affects 
the ion
correlations and the ion condensation on DNA surface, which are
important contributions to the electrostatic
potential around the DNA molecule \cite{gavryushov,stigter1996}.

\begin{figure}
    \epsfxsize=15cm 
    \epsfysize=15cm 
 ~\hfill\epsfbox{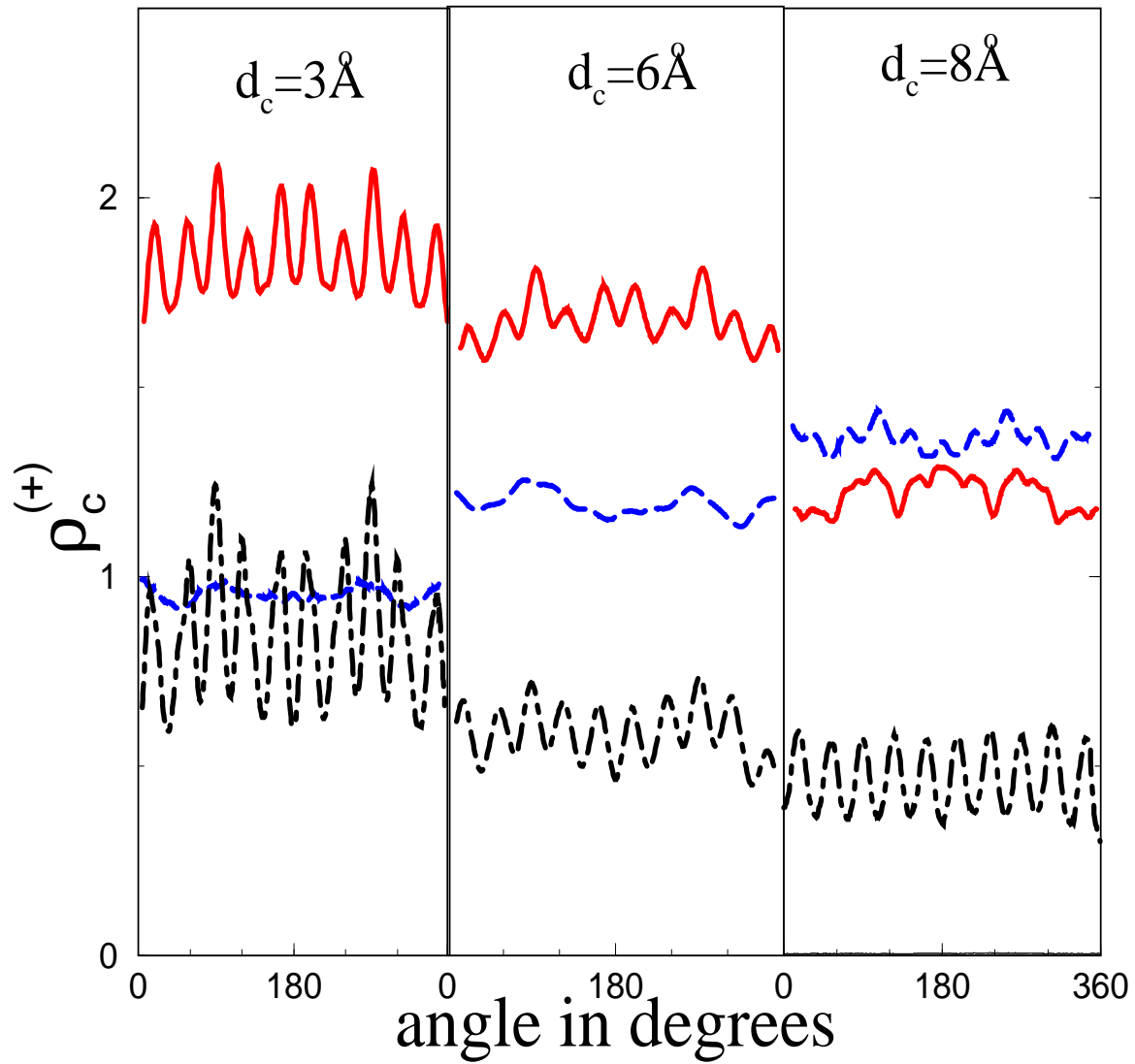}\hfill~
 \caption{Panoramic view of  cation  density near DNA surface for
   $q_c=1$, $q_s=1$ (Set 1), $C_s=$0.1 Mol/l and the MAM. Different cation diameters, from left
   to right: $d_c=3$\AA,\,  $d_c=6$\AA, \, $d_c=8$\AA. Full line- cation
   distribution in the minor groove, dashed
 line- cation distribution in the major groove, dot-dashed line-
cation distribution on the phosphate strands. Note that the the cation
   adsorption in the major groove exceeds the cation adsorption in the
   minor groove for $d_c=8$\AA, collate the full and dashed lines in the
   rigth side of figure.}
 \label{fig16}
\end{figure}

 We vary the hydrodynamic radii of solvated ions between 3\AA \, and
8\AA \, \cite{conrad1988biopol,conway1981}. The simulation results for
three different
ion diameters, $d_c$=3\AA, 6\AA, 8\AA \, are shown in Figure
\ref{fig16}. The gap between the cationic groove occupations decreases as 
the ion radius increases. Furthermore, more cations go to the major groove
whereas the minor groove and the strand ionic occupations shrink. In total, 
a small amount of cations and anions
condense on the DNA surface. Nevertheless all qualitative findings obtained for
ions with diameter $d_c$=3\AA \, and outlined in the previous 
section, retain
valid for twice enlarged ion size $d_c$=6\AA. 
 A further increase of the ionic size up to $d_c$=8\AA , however, makes the ion 
intrusion
into the minor groove a very rare event. The major groove cationic
charge then exceeds the minor groove cationic charge, see the right-hand
side of Figure \ref{fig16}. Thus  the salt ions "physically"
can not explore the full details of the elaborated DNA model. In other
words, the salt ions start to "feel" the MAM shape as a less
elaborated DNA model, like the ECM. Thus the simulation results for
the MAM with $d_c$=8\AA \, qualitatively resemble the results for the ECM 
with $d_c$=6\AA \, or $d_c$=3\AA.

\section{conclusions}
 In this work we have studied several models of a DNA polyelectrolyte
system containing a mixture of mono- and multivalent ions
 within the framework of a continuum
dielectric approach. We have neglected the granular nature of
water and concentrated only on electrostatics of the ion condensation on
the DNA surface. Summarizing the
results obtained, we have first shown that the small ion condensation 
pattern on
the DNA surface strongly depends on the geometry of the DNA model
used. While in the simple cylindrical model
 cations predominantly bind to the phosphate strands, 
in the more realistic  Montoro-Abascal
model the minor
groove becomes the principal site of cationic binding. Our simulation
results also indicate that the anion condensation is less sensitive to
the DNA model shape. We have further investigated the occupancy and 
charging of
DNA grooves as function of increasing cation valency.

 We find that the adsorbed ion pattern change with
increasing counterion valency is as follows:\\  
i) the cations leave both the major and minor grooves to
 the phosphate strands, \\
ii) there are more cations in the minor groove  than in the major groove
exhibiting a structuring 
reminiscent to the spine of solvation,\\
iii) the accumulated cationic charges in the major groove are almost 
independent of the counterion valency.

An increase in salt concentration leads to the
following effects:\\
for monovalent salt,\\
i) the {\it major} groove keeps its total charge at a constant value,\\
ii) there is a constant cationic charge asymmetry between the DNA
grooves,\\
for multivalent salt,\\
iii) the {\it minor} groove keeps its total charge at a constant value,\\
iv) the major groove is getting neutralized and a DNA overcharging occurs.

All these trends represent important information for the implementation of
more phenomenological theories for the interaction between two DNA molecules
 such as the Kornyshev-Leikin theory \cite{kornyshevandleikin}
where the number of condensed ions on the phosphate strands and on the grooves
is a key  input quantity. As our results indicate, however, one cannot 
assume constant fractions of adsorbed ions when the salt concentration
and/or the counterion valency is varied.

Let us finally discuss some improvements of our model which could be done step by step
in a more realistic description of DNA.
First we assume that the persistence length of DNA does not depend
from the added salt concentration in order to fit our set up of
infinitely stiff single DNA
\cite{williams,lu2002,ariel2001,duguid1995,sottas1999bj,khan1999,baumann1997,kunze2002,liu2000jcp,widom,bloomfield1997,ramin1999}.
The effect of added salt on the DNA stifness can be taken into account
only in simulations with finite DNA fragments. 
Second, refinement of the
present models may account for the specific short-range ion-DNA
interactions, or specific "bonding" of ions to the
 DNA surface,  on the basis of an effective ion-ion and ion-DNA 
interactions
\cite{gulbrand1989,pettitt1986jcp}.  
Such chemisorption is believed to contribute to the force-angle 
dependence, which could be lasted over distances larger the Debye
length, 15\AA-30\AA \,
\cite{kornyshevandleikin,holgerprl2002,cherstvy2002jpc}.
Third, and maybe most important of all, the solvent granularity and
the space-dependent dielectric constant  were disregarded completely
\cite{yang1995bj,lyubartsev1995,Pack1993,jaya2,conrad1988biopol,kusalik1988,gilson1985,jaya1,jayaram1998jpc,mazur,petsev2000bj,luka1,maarel1999,messinadielectric}.
Fortunately, in most applications, the dielectric discontinuity and
dielectric saturation effects have canceling influences in such
properties as the energy or the number of condensed ions. That is why the
use of a homogeneous solvent with a constant $\epsilon$ is believed to be
not as crude as it seems from the first point of view
\cite{simulationswithmolecularwater,lyubartsev1995pre}.  
Therefore we believe that the general
trends found in our study will be stable with respect to a 
more realistic description of the solvent.

The other experimentally inspired issue is the mechanics of B-DNA which 
allows 
the minor groove to open and close to
accommodate a divalent cations \cite{chiu2000jmb}. The work to account
for this effect as well as the ion chemisorption in the major groove is an
objective of future work.  

\acknowledgments
E. A. thanks R. Podgornik for fruitful discussions of some results of
this paper. We thank the DFG for financial support.

\begin{table}
\caption{Parameters used for the different simulation sets.}
\begin{tabular}{lcc} 
Set & $q_c$ & $q_s$ \\ 
  Set 1 &   1  &  1 \\ 
  Set 2 &   1  &  2 \\
  Set 3 &   3  &  2 \\ 
  Set 4 &   2  &  1 \\
  Set 5 &   3  &  1 \\ 
  Set 6 &   2  &  2 \\
\tableline
   
\end{tabular} 
\label{tab2}
\end{table}

%------------------------figure 1---------------------------------

 \newpage
% ----------------------------figure 2------------------------------

  \newpage
%----------------------------figure 3-----------------------------

 \newpage
%-----------------------------figure 4-------------------------

  \newpage
%------------------------figure 5---------------------------------

  \newpage

%------------------------figure 6---------------------------------

  \newpage
%------------------------figure 7--------------------------------

  \newpage
%------------------------figure 8---------------------------------

  \newpage
%------------------------figure 9---------------------------------

 \newpage
%------------------------figure 10---------------------------------

 \newpage
%------------------------figure 11---------------------------------

  \newpage
%------------------------figure 12---------------------------------

 \newpage
%------------------------figure 13---------------------------------

 \newpage
%------------------------figure 14---------------------------------

\end{document}